\documentclass[twocolumn,apj,numberedappendix]{openjournal}

\usepackage[T1]{fontenc}
\usepackage{amsmath,amsfonts,mathrsfs}
\usepackage{graphicx}
\usepackage{placeins}

\usepackage{bbm}
\usepackage{bm}
\usepackage{comment}
\usepackage{booktabs}
\usepackage{color}
\usepackage[hidelinks]{hyperref}

\makeatletter
\@ifundefined{AddToHook}{}{%
  \let\OJAp@putlastpage\lastpage@putlabel
  \let\lastpage@putlabel\relax
  \AddToHook{enddocument/afterlastpage}{\OJAp@putlastpage}%
}
\makeatother

\def\deltadir{\delta_{\rm D}}

\def\cov{\mathbbm{Cov}}

\def\pd{{\rm d}}

\hypersetup{
  pdftitle={Selection--Clustering Coupling and Pair-Midpoint Anisotropy from Observer Motion},
  pdfauthor={Tsutomu T. Takeuchi},
  pdfsubject={Pair-conditioned selection and observer-induced anisotropy of cosmological source-catalog correlations},
  pdfkeywords={observer motion, selection-clustering coupling, pair-conditioned selection, pair-midpoint anisotropy, velocity estimator}
}

\begin{document}
\emergencystretch=2em

\submitted{}
\journalinfo{}
\shorttitle{Selection--Clustering Coupling from Observer Motion}
\shortauthors{T. T. Takeuchi}
\title{Selection--Clustering Coupling and Pair-Midpoint Anisotropy from Observer Motion}
\author{Tsutomu T. Takeuchi}
\email[Corresponding author: ]{tsutomu.takeuchi.ttt@gmail.com}
\thanks{ORCID: \href{https://orcid.org/0000-0001-8416-7673}{0000-0001-8416-7673}.}
\affiliation{Division of Particle and Astrophysical Science, Nagoya University, Furo-cho, Chikusa-ku, Nagoya 464--8602, Japan}
\affiliation{The Research Center for Statistical Machine Learning, The Institute of Statistical Mathematics, 10-3 Midori-cho, Tachikawa, Tokyo 190--8562, Japan}

\begin{abstract}
Observer motion couples source selection, clustering, and estimator normalization in cosmological pair statistics. Existing calculations distinguish the source-count dipole from the modulation of clustering, including selection-dependent bias. We develop an observable-level formulation that retains arbitrary marked pair kernels, nonfactorizable pair inclusion, and explicit reference measures. Its independent-selection limit expresses the residual as a covariance of the selection score with pair excess; in a rank-one bias model this is the logarithmic response of the selected mean bias and recovers the corresponding bias-derivative description. Two populations can have the same count-dipole coefficient and isotropic correlation but opposite two-point boost responses. We show that local division by endpoint means is not equivalent to a scalar normalization of integrated pair counts, and derive the contact terms and self-pair subtraction required for discrete harmonic estimators. Exchange symmetry permits pair-midpoint anisotropy but forbids an ordinary unordered same-tracer separation dipole. The known neighboring-multipole templates are recovered, with explicit conditions for velocity--selection degeneracy and its uncertainty cost. Finite-boost forward models independently test the response, including a nonfactorizable retained-pair example. The result is a common framework for bias-response limits, general pair selection, normalization, and covariance conventions; the numerical illustrations are controlled models rather than survey forecasts. An observational route is specified through radio and infrared catalog selections, full-pipeline response calibration, and common-velocity consistency tests.
\end{abstract}



\section{Introduction}
\label{sec:introduction}

Observer motion changes cosmological source catalogs through aberration and Doppler modulation. The one-point number-count dipole is described by the Ellis--Baldwin calculation and its extensions to realistic populations and selection functions \citep{1984MNRAS.206..377E,2026ApJS..285....3T}. Its extension to two-point observables requires more than the mean count response: one must specify the selected pair measure, the statistic applied to that measure, and the normalization of the measured field or pair count.

This problem is relevant to the cosmic dipole anomaly. Radio and mid-infrared catalog analyses have reported dipoles broadly aligned with the CMB direction but with amplitudes exceeding the simplest CMB-based prediction \citep{2012MNRAS.427.1994G,2017MNRAS.471.1045C,2021ApJ...908L..51S,2022ApJ...937L..31S,2023MNRAS.525..231D,2024MNRAS.531.4545O,2025A&A...697A.112W}. Clustering and catalog-systematics analyses provide complementary information \citep{2023ApJ...943..116T,2026MNRAS.546ag201V,2026MNRAS.546ag248O}; see \citet{2025RvMP...97d1001S} for a review. A discrepancy in a count dipole does not alone identify its origin. Population properties, selection, nearby structure, calibration, and survey geometry also enter its interpretation.

Population-dependent clustering modulation has an established observer-motion description. \citet{2024JCAP...06..045L}, building on observer-motion studies including \citet{2022MNRAS.512.3895D,2023PhRvD.107j3514D}, distinguish the count-dipole coefficient $D_{\rm kin}$ from the fluctuation coefficient $F_{\rm kin}$. Their calculation includes flux-dependent bias, joint velocity--modulation estimation, power-law degeneracies, and a nonlinear cross-spectrum extension. These results provide the benchmarks and restricted limits of the present formulation.

Our objective is to express the response at the level of a specified selected marked-pair observable, without requiring a single flux threshold, a single effective bias, or a map-factorized pair prescription. This formulation separates three logically different operations: changing the population represented by a catalog, changing the geometrical locations of its pairs, and changing the reference against which those pairs are normalized. It also distinguishes the covariance of a source map from the expectation of a separately retained or weighted pair list. The latter need not be the covariance of any single map.

For independent objectwise selection, the selected one-point mark distribution is reweighted when one conditions on an actual pair. A smooth selection perturbation therefore yields a covariance between its score and the marked pair-density ratio. The normalized-measure differentiation identity makes this dependence explicit. In a rank-one linear-bias model it gives
\begin{align}
\mathcal M_{\rm sel}
=\frac{\cov(s,b)}{\bar b}
=\left.\frac{\partial\ln\bar b_\varepsilon}{\partial\varepsilon}\right|_0,
\label{eq:intro_bias_response}
\end{align}
where $\varepsilon=\ln\delta$, $s$ is the selection score, and $\bar b_\varepsilon$ is the mean bias of the perturbed selected population. For a common flux response $q$ and a single threshold, this becomes $-q\,\partial\ln\bar b/\partial\ln S_{\rm lim}$. The covariance representation thus recovers a selection-dependent bias response while also identifying the joint mark information needed to evaluate it. Section~\ref{subsec:prior_work_correspondence} makes the comparison to redshift-resolved number counts explicit.

The corresponding locally normalized same-tracer angular correlation has an exchange-even midpoint response,
\begin{align}
\mathcal R_{aa}^{\rm mid}(\theta)
={}&2\sin\left(\frac{\theta}{2}\right)\frac{\pd w}{\pd\theta}
+2\cos\left(\frac{\theta}{2}\right)\mathcal M_{\rm sel}w(\theta).
\label{eq:intro_worked_midpoint_response}
\end{align}
Here $w=\bar b^2w_{\rm m}$ for the rank-one model. The coefficient $\mathcal M_{\rm sel}$ does not follow from the one-point dipole coefficient alone. We exhibit two admissible populations with the same selected density, count-dipole response, and isotropic angular correlation, but opposite $\mathcal M_{\rm sel}$. We also give a finite-boost benchmark with nonfactorizable pair retention, where a single effective-bias coefficient is insufficient.

Normalization is essential to the interpretation of Eq.~\eqref{eq:intro_worked_midpoint_response}. Dividing a covariance locally by the two endpoint means is not equivalent to dividing an integrated signed pair count by a single random-pair count. The latter can retain the one-point intensity modulation even when its random catalog follows the boosted mean. Likewise, self-pairs contribute to a raw harmonic product but not to a factorial pair statistic. We derive the required kernel reweighting and self-pair subtraction, and keep the mean response of discreteness noise separate from its contribution to estimator variance.

Endpoint exchange supplies a further exact constraint. An unordered same-tracer pair cannot support a kernel odd solely under endpoint reversal. An observer boost instead produces an allowed pair-midpoint anisotropy, represented by $L=1$ bipolar spherical-harmonic (BipoSH) coefficients and neighboring ordinary multipoles. This is distinct from relativistic dipoles of ordered cross-correlations. Relativistic observed number counts contain density, redshift-space, Doppler, potential, lensing, and evolution terms \citep{2010PhRvD..82h3508Y,2011PhRvD..84f3505B,2014JCAP...09..037B,2014JCAP...11..013B}; tracer or radial ordering permits separation-odd contributions \citep{2014PhRvD..89h3535B,2018PDU....19..109R}. Standard redshift-space anisotropy from source peculiar velocities is another distinct contribution \citep{1987MNRAS.227....1K,1998ASSL..231..185H,2004PhRvD..70h3007S}. Our angular construction does not replace a full relativistic light-cone calculation.

The contribution developed below is an observable-level extension and calibration framework. We give the complete boost derivative of a selected pair functional, including nonfactorizable pair selection and explicit analysis kernels; derive its locally normalized and integrated forms; connect smooth selection scores and sharp-boundary responses to effective-bias derivatives; and specify the discrete-map and noise conventions needed for harmonic inference. Known aberration--modulation templates and their degeneracies are recovered as special cases. Reproducible forward-model tests assess the response formulas independently of their analytic decomposition. The numerical examples are validation models, not forecasts for a named survey.

The observational use of this framework is to compare a measured pair-midpoint projection or neighboring-multipole statistic with the prediction for the same catalog construction. Radio and mid-infrared samples already used for count-dipole and clustering analyses provide concrete targets, while flux and color variations can test the additional selection response. Section~\ref{sec:observational_connection} specifies the required catalog inputs, a finite-boost calibration of the complete measurement pipeline, and consistency tests across selections. These applications do not require an ordinary same-tracer separation dipole, and they do not assume that the precision of the controlled examples is the precision of an observed survey.

Section~\ref{sec:one_point_boost} fixes the one-point conventions. Section~\ref{sec:pair_observables_exchange} defines the selected pair measures and exchange rules. Section~\ref{sec:two_point_boost_response} derives the general response, its score representation, and its relation to bias-response calculations. Section~\ref{sec:normalization_contact} distinguishes normalization and contact-term conventions. Section~\ref{sec:harmonic_coupling} gives harmonic inference and identifiability. Section~\ref{sec:validation} specifies the independent numerical benchmarks. Section~\ref{sec:observational_connection} connects the calculation to catalog measurements and calibration tests, and Section~\ref{sec:conclusion} summarizes the scope. Appendices~\ref{app:selection_response}--\ref{app:factorial_covariance} provide boundary, geometrical, harmonic, and covariance details.

\section{One-Point Boost as the Endpoint Baseline}
\label{sec:one_point_boost}

We summarize only the one-point ingredients needed below. Primes denote cosmic-rest-frame directions and radiometric quantities, whereas unprimed quantities are measured by an observer moving with velocity $\bm v$. Define
\begin{align}
\bm\beta
&\equiv
\frac{\bm v}{c},
\qquad
\beta
\equiv
|\bm\beta|,
\qquad
\gamma
\equiv
\left(1-\beta^2\right)^{-1/2}.
\end{align}
For an observed direction $\hat{\bm n}$, the Doppler factor and first-order aberration are
\begin{align}
\delta(\hat{\bm n})
&\equiv
\frac{1}{\gamma\left(1-\bm\beta\cdot\hat{\bm n}\right)},
\label{eq:doppler_factor}
\\
\hat{\bm n}'
&=
\hat{\bm n}
-
\bm\beta_{\perp}(\hat{\bm n})
+
\mathcal O(\beta^2),
\nonumber\\*
\bm\beta_{\perp}(\hat{\bm n})
&\equiv
\bm\beta
-
\left(\bm\beta\cdot\hat{\bm n}\right)
\hat{\bm n}.
\label{eq:first_order_aberration}
\end{align}
The local infinitesimal parameter is
\begin{align}
\epsilon(\hat{\bm n})
\equiv
\ln\delta(\hat{\bm n})
=
\bm\beta\cdot\hat{\bm n}
+
\mathcal O(\beta^2),
\label{eq:epsilon_definition}
\end{align}
and the solid-angle Jacobian is
\begin{align}
\frac{\pd\Omega'}{\pd\Omega}
=
\delta^2(\hat{\bm n})
=
1+2\epsilon(\hat{\bm n})
+
\mathcal O(\beta^2).
\label{eq:solid_angle_jacobian}
\end{align}

The local-score formulas below assume catalog marks whose specified transformation at fixed observed direction closes on the local Doppler factor. Resolved tensor marks or instrument-orientation responses may require a more general boost derivative; in that case retain the full pipeline derivative rather than identifying it with a scalar Doppler score. Let $\bm y$ denote the source marks used to construct the catalog, including fluxes, colors, redshift information, morphology summaries, or classification scores. The observer-frame marks $\bm y_\delta$ define the infinitesimal generator
\begin{align}
\mathcal B_{\bm y}
\equiv
\sum_A
\left.
\frac{\partial y_\delta^A}{\partial\ln\delta}
\right|_{\delta=1}
\frac{\partial}{\partial y^A}.
\label{eq:mark_boost_generator}
\end{align}
For completeness, the unresolved-source radiometric transformation is $S_\nu(\nu)=\delta S_{\nu'}'(\nu/\delta)$. If $\mathcal T_{\rm b}(\nu)$ is a fixed instrumental band response, including the frequency weighting used in the measurement, then
\begin{align}
S_{{\rm b},\delta}
&=\delta^2\int_0^\infty\mathcal T_{\rm b}(\delta\nu')S_{\nu'}'(\nu')\,\pd\nu',\label{eq:finite_band_flux_r4}\\
q_{\rm b}\equiv\left.\partial_{\ln\delta}\ln S_{{\rm b},\delta}\right|_1
&=2+\frac{\int_0^\infty\nu'\mathcal T_{\rm b}'(\nu')S_{\nu'}'(\nu')\,\pd\nu'}{\int_0^\infty\mathcal T_{\rm b}(\nu')S_{\nu'}'(\nu')\,\pd\nu'}.
\label{eq:finite_band_response_r4}
\end{align}
These equations follow from invariance of $I_\nu/\nu^3$, aberration of the unresolved source solid angle, and change of frequency variable. Integration by parts gives $q_{\rm b}=1+\alpha$ for $S_{\nu'}'\propto\nu'^{-\alpha}$ when the band-boundary term vanishes. A general spectrum or bandpass produces a source-dependent response \citep{2026ApJS..285....3T}. Reference redshift and a magnitude color obey
\begin{align}
1+z_\delta&=\frac{1+z}{\delta},\\
\left.\partial_{\ln\delta}c_{{\rm a}{\rm b},\delta}\right|_1
&=-\frac{2.5}{\ln10}(q_{\rm a}-q_{\rm b}),
\label{eq:representative_mark_responses_r4}
\end{align}
where $c_{{\rm a}{\rm b}}=-2.5\log_{10}(S_{\rm a}/S_{\rm b})$. Derived classification variables must be transformed through their actual measurement pipeline. Intrinsic labels that do not change under a boost have zero generator component. We write
\begin{align}
\dot{\mathcal W}(\bm y,\hat{\bm n})
\equiv\left.\partial_{\ln\delta}\mathcal W(\bm y_\delta,\hat{\bm n})\right|_1
=\mathcal B_{\bm y}\mathcal W(\bm y,\hat{\bm n}).
\label{eq:selection_derivative_r4}
\end{align}
A stochastic pipeline can have a more general inclusion-law derivative, which replaces the final generator expression without changing the integral identities below. Appendix~\ref{app:selection_response} gives its boundary and forward-model forms.

Let $\rho_1'(\hat{\bm n}',\bm y)$ be the parent first factorial-moment density and let $0\leq\mathcal W(\bm y,\hat{\bm n})\leq1$ denote the catalog inclusion probability. Number conservation gives the selected intensity per observed solid angle,
\begin{align}
n_{\bm\beta}(\hat{\bm n})
={}&
\delta^2(\hat{\bm n})
\int
\rho_1'
\left(
\hat{\bm n}'(\hat{\bm n}),
\bm y
\right)
\mathcal W
\left(
\bm y_\delta,
\hat{\bm n}
\right)
\,\pd\bm y.
\label{eq:general_selected_intensity}
\end{align}
For a statistically isotropic parent population, $\rho_1'(\hat{\bm n}',\bm y)=f(\bm y)$, so
\begin{align}
n_{\bm\beta}(\hat{\bm n})
=
\delta^2(\hat{\bm n})
\int
f(\bm y)
\mathcal W
\left(
\bm y_\delta,
\hat{\bm n}
\right)
\,\pd\bm y.
\label{eq:isotropic_selected_intensity}
\end{align}
At fixed observed direction, the generalized one-point coefficient is
\begin{align}
\mathcal A[\mathcal W,f](\hat{\bm n})
&\equiv
\left.
\frac{\partial\ln n_\delta(\hat{\bm n})}
{\partial\ln\delta}
\right|_{\delta=1}
\nonumber\\*
&=
2+
\frac{
\displaystyle
\int
f(\bm y)
\mathcal B_{\bm y}\mathcal W(\bm y,\hat{\bm n})
\,\pd\bm y
}{
\displaystyle
\int
f(\bm y)
\mathcal W(\bm y,\hat{\bm n})
\,\pd\bm y
}.
\label{eq:generalized_EB_functional}
\end{align}
The first term is the solid-angle contribution. The second is the selected-population response at the catalog boundary or its smooth probabilistic analogue. For uniform angular selection,
\begin{align}
n_{\bm\beta}(\hat{\bm n})
=
\bar n_0
\left[
1+
\mathcal A\,
\bm\beta\cdot\hat{\bm n}
\right]
+
\mathcal O(\beta^2).
\end{align}

For a sharp flux threshold, write the parent marks as $(S,\bm\eta)$ and let $q=\left.\partial_{\ln\delta}\ln S_\delta\right|_1$. Defining
\begin{align}
f_S(S)&=\int f(S,\bm\eta)\,\pd\bm\eta,\\
n_0(>S)&=\int_S^\infty f_S(t)\,\pd t,\\
x(S_{\rm lim})&=-\left.\partial_{\ln S}\ln n_0(>S)\right|_{S_{\rm lim}}\nonumber\\*
&=\frac{S_{\rm lim}f_S(S_{\rm lim})}{n_0(>S_{\rm lim})},\\
\langle q\rangle_{\partial\mathcal W}
&=\frac{\int f(S_{\rm lim},\bm\eta)q(S_{\rm lim},\bm\eta)\,\pd\bm\eta}{f_S(S_{\rm lim})}.
\label{eq:boundary_slope_definitions_r4}
\end{align}
the derivative $\dot{\mathcal W}=qS\delta_{\rm D}(S-S_{\rm lim})$ reduces Eq.~\eqref{eq:generalized_EB_functional} to
\begin{align}
\mathcal A
=
2+
x(S_{\rm lim})
\left\langle q\right\rangle_{\partial\mathcal W}.
\label{eq:threshold_generalized_EB}
\end{align}
If the sources share a power-law spectrum, $q=1+\alpha$, then
\begin{align}
\mathcal A_{\rm EB}
=
2+x(S_{\rm lim})(1+\alpha),
\label{eq:classical_EB_coefficient}
\end{align}
which recovers the Ellis--Baldwin result \citep{1984MNRAS.206..377E}. Only the local count slope is needed at first order; an exact finite-boost power law additionally requires power-law counts over the interval swept by the threshold. If $\mathcal A$ or $n_0$ varies on the sky, their product with $\bm\beta\cdot\hat{\bm n}$ is not in general a pure dipole. For uniform full-sky selection the first-order perturbation of the total count integrates to zero, although the directional intensity has a nonzero dipole. The coefficient fixes the one-point intensity response, not the dependence of a normalized pair statistic on joint marks and its reference measure.

\section{Pair Observables and Exchange Constraints}
\label{sec:pair_observables_exchange}

We use the standard factorial-moment conventions for marked point processes \citep{2003Daley_point_processI,Last_Penrose_2017}. At zero boost, a state is $Z=(\hat{\bm n},\bm y)$ with measure $\pd Z=\pd\Omega\,\pd\bm y$; sums replace integrations over discrete marks. The parent densities $\rho_{1,a}'$ and $\rho_{2,ab}'$ describe single objects and ordered distinct pairs. The identical-object diagonal is excluded even when tracer samples overlap. Factorial densities contain no self-pair contact term, but this exclusion does not remove sampling noise from an estimator.

\subsection{Object inclusion and effective pair selection}
\label{subsec:pair_selection_probabilities}

For independent objectwise inclusion, the selected intensity and second density are $\lambda_a=\rho_{1,a}'\mathcal W_a$ and $\rho_{2,ab}'\mathcal W_a\mathcal W_b$. A correlated detection mechanism instead requires conditional single and joint inclusion probabilities. Denoting their effective pair factor by $\mathcal C_{ab}$, write
\begin{align}
\lambda_a(Z)&=\rho_{1,a}'(Z)\mathcal W_a(Z),\\
\lambda_{ab}^{(2)}(Z_1,Z_2)
&=\rho_{2,ab}'(Z_1,Z_2)\mathcal W_a(Z_1)\mathcal W_b(Z_2)\nonumber\\*
&\quad\times\mathcal C_{ab}(Z_1,Z_2).
\label{eq:selected_second_marked_density}
\end{align}
When these are factorial densities of an actual detected point process, the inclusion laws must be mutually consistent. An arbitrary nonnegative $\mathcal C_{ab}$ cannot simply be declared to define such a process. In particular, $\mathcal C_{ab}$ is a ratio of joint to product inclusion probabilities and need not be bounded by one. If single and joint inclusion probabilities are defined under the same conditioning, they obey the usual bounds $\max(0,\pi_1+\pi_2-1)\leq\pi_{12}\leq\min(\pi_1,\pi_2)$; a global detection model imposes additional consistency conditions.

There is a second, distinct use of a pair factor. An independently selected object catalog may be followed by a pair-retention rule $R_{ij}$, applied to its pair list without removing objects. If
\begin{align}
\mathbb E[R_{ij}\mid Z_i,Z_j,\text{objects selected}]
=\mathcal C_{ab}(Z_i,Z_j),
\end{align}
then $0\leq\mathcal C_{ab}\leq1$ for Bernoulli retention, and the right-hand side of Eq.~\eqref{eq:selected_second_marked_density} is the effective retained-pair measure. It is not generally the second factorial-moment measure of a new source catalog. We use the same notation for this effective measure when calculating pair expectations, but do not identify it with a map covariance. The explicit validation model in Section~\ref{sec:validation} uses this pair-list interpretation.

In either case, assume exchange consistency:
\begin{align}
\mathcal C_{ab}(Z_1,Z_2)&=\mathcal C_{ba}(Z_2,Z_1),
\label{eq:pair_completeness_exchange}\\
\lambda_{ab}^{(2)}(Z_1,Z_2)&=\lambda_{ba}^{(2)}(Z_2,Z_1),
\label{eq:cross_exchange_identity}\\
\lambda_{aa}^{(2)}(Z_1,Z_2)&=\lambda_{aa}^{(2)}(Z_2,Z_1).
\label{eq:auto_exchange_identity}
\end{align}
For an actual point process, the second relation follows by relabeling objects. For an additionally retained pair list, the same identity follows when the retention is assigned to unordered pairs or is symmetrized. Inverse-completeness analysis weights belong in the kernel; they do not remove the detection probability from the expectation. A probability factor and its correcting weight must each be represented once.

\subsection{Kernels, references, and population targets}
\label{subsec:kernelized_pair_targets}

For objectwise selection followed by optional pair retention, the ordered pair statistic has expectation
\begin{align}
P_{ab}[K]
&\equiv\mathbb E\left[\sum_{i\in X_a}\sum_{j\in X_b}'R_{ij}K_{ab}(Z_i,Z_j)\right]
\nonumber\\*
&=\int K_{ab}(Z_1,Z_2)\lambda_{ab}^{(2)}(Z_1,Z_2)\,\pd Z_1\pd Z_2.
\label{eq:expected_kernelized_pair_statistic}
\end{align}
Set $R_{ij}=1$ for a catalog whose joint detection is already included in $\lambda^{(2)}$. The kernel specifies separation bins, marks, ordering, weights, or a harmonic projection. It may be complex for harmonic coefficients. The product reference and its connected difference are
\begin{align}
M_{ab}[K]&=\int K_{ab}\lambda_a\lambda_b\,\pd Z_1\pd Z_2,
\label{eq:product_reference_definition}\\
C_{ab}[K]&=P_{ab}[K]-M_{ab}[K].
\label{eq:connected_functional_definition}
\end{align}
For an actual point process this is a factorial connected moment. For a separately retained pair list it is an operational excess over the product reference and may include pair-retention structure even when the source process is Poisson.

Given a nonnegative $K_0$ with $M_{ab}[K_0]>0$, define
\begin{align}
\Xi_{ab}[K\mid K_0]=\frac{C_{ab}[K]}{M_{ab}[K_0]}.
\label{eq:signed_pair_projection}
\end{align}
This ratio of population integrals is not an integral of a locally normalized correlation. Neither is it generally the exact expectation of a finite-sample ratio. Finite random catalogs, estimated monopoles, and data-dependent weights require their own sampling treatment. We return to the distinction in Section~\ref{sec:normalization_contact}.

For independent random catalogs with prescribed intensities $r_a,r_b$, define $DD=P$, $DR=\int K\lambda_a r_b$, $RD=\int Kr_a\lambda_b$, and $RR=\int Kr_ar_b$. The population Landy--Szalay numerator \citep{1993ApJ...412...64L} satisfies
\begin{align}
L_{ab}[K;r]&=DD-DR-RD+RR\nonumber\\*
&=C_{ab}[K]\nonumber\\*
&\quad+\int K_{ab}(\lambda_a-r_a)(\lambda_b-r_b)\,\pd Z_1\pd Z_2.
\label{eq:LS_reference_mismatch}
\end{align}
Here all four terms use the stated kernel; $DD$ contains any actual pair loss. If the random pairs are instead subjected to an additional modeled retention law, that changes the reference and must be included explicitly, not silently absorbed into this identity. When $r_a=\lambda_{a,0}$ at the unboosted point, the explicit mismatch is second order in a perturbation, but $C[K]$ can still retain first-order one-point modulation.

\subsection{Exchange theorem and admissible angular projections}
\label{subsec:exact_exchange_theorem}

Define the label-aware transpose by $(K^\top)_{ab}(Z_1,Z_2)=K_{ba}(Z_2,Z_1)$. Changing integration variables gives
\begin{align}
P_{ab}[K]=P_{ba}[K^\top],\qquad
C_{ab}[K]=C_{ba}[K^\top].
\end{align}
For an auto-kernel $K_-^\top=-K_-$, this implies
\begin{align}
P_{aa}[K_-]=M_{aa}[K_-]=C_{aa}[K_-]=0.
\label{eq:auto_odd_kernel_expectation_zero}
\end{align}
The finite ordered pair sum also vanishes: contributions from $(i,j)$ and $(j,i)$ cancel if their selection and retention are identical. A common mask or depth variation does not change this pairing. For an unordered pair list the kernel must be symmetrized; orienting an otherwise identical pair by an arbitrary catalog index does not define a physical separation dipole.

For an odd kernel, $DD=RR=0$ and $RD=-DR$. The full Landy--Szalay numerator therefore vanishes. Its familiar reduction $DD-2DR+RR$ assumes an exchange-even auto-kernel and cannot be applied to an odd projection.

Away from coincident and antipodal configurations, define
\begin{align}
u_{12}&=\hat{\bm n}_1\cdot\hat{\bm n}_2,\qquad \theta_{12}=\arccos u_{12},\\
\hat{\bm N}_{12}&=\frac{\hat{\bm n}_1+\hat{\bm n}_2}{\sqrt{2(1+u_{12})}},\qquad
\hat{\bm s}_{12}=\frac{\hat{\bm n}_1-\hat{\bm n}_2}{\sqrt{2(1-u_{12})}}.
\label{eq:angular_pair_geometry}
\end{align}
The midpoint direction is exchange even and the oriented separation is odd. For an isotropic, parity-even scalar reference problem, a correction linear in a single boost can depend on $\bm\beta\cdot(\hat{\bm n}_1+\hat{\bm n}_2)$ and $\bm\beta\cdot(\hat{\bm n}_1-\hat{\bm n}_2)$, with coefficients depending on $u_{12}$. Auto-pair symmetry removes the second coefficient. This elementary symmetry constraint is a consistency requirement on the observable, not a new relativistic interaction.

\subsection{Marked and cross-tracer ordering}
\label{subsec:marked_auto_pair_ordering}

A separation-odd factor can appear in a legitimate total kernel if another factor is endpoint odd. For example,
\begin{align}
K_0(Z_1,Z_2)\operatorname{sgn}[m(\bm y_1)-m(\bm y_2)]\,\hat{\bm q}\cdot\hat{\bm s}_{12}
\end{align}
is exchange even for symmetric $K_0$. It is a mark-ordered statistic, not an ordinary unordered separation dipole. Distinct tracers or explicit radial ordering provide other admissible endpoint labels. The joint mark--location interpretation is naturally expressed by marked point processes \citep{2026MNRAS.549g1015T}. Relativistic ordered cross-correlation dipoles occupy this labeled-pair sector and remain separate from the same-tracer midpoint response.

\section{Pair-Conditioned Selection and the Boost Response}
\label{sec:two_point_boost_response}

\subsection{Exact pair functional and regularity assumptions}
\label{subsec:exact_pair_boost}

Vary the observer velocity along $t\bm\beta$ and define
\begin{align}
\delta_{\bm\beta}F\equiv\left.\frac{\pd F_{t\bm\beta}}{\pd t}\right|_{t=0}.
\label{eq:directional_boost_variation}
\end{align}
This is linear in $\bm\beta$. In an observed-coordinate representation, the directions $\hat{\bm n}_i$ are held fixed while $\bm y_i$ are fixed parent-mark coordinates. Define $\pd\Gamma_{12}=\pd\Omega_1\pd\Omega_2\pd\bm y_1\pd\bm y_2$ and $\epsilon_i=\bm\beta\cdot\hat{\bm n}_i$. The exact selected pair functional is
\begin{align}
P_{ab,\bm\beta}[K_{\bm\beta}]
={}&\int \delta_1^2\delta_2^2 K_{ab,\bm\beta}
\rho_{2,ab}'(\hat{\bm n}'_1,\bm y_1;\hat{\bm n}'_2,\bm y_2)
\nonumber\\*
&\times\mathcal W_a(\bm y_{1,\delta_1},\hat{\bm n}_1)
\mathcal W_b(\bm y_{2,\delta_2},\hat{\bm n}_2)
\mathcal C_{ab,\bm\beta}\,\pd\Gamma_{12}.
\label{eq:exact_boosted_pair_functional}
\end{align}
The kernel and pair factor are pulled back to these same variables. They may depend on transformed marks or on an explicit analysis prescription. A bin defined at fixed observed angular separation has no independent angular derivative here. Masks and instrumental depth are also fixed functions of the observed directions. The reference marks are integration labels, so Eq.~\eqref{eq:exact_boosted_pair_functional} contains no additional mark-space Jacobian. An equivalent expression in observer-frame marks includes the appropriate Jacobian by change of variables.

We assume finite factorial moments over the analysis window and sufficient integrability to differentiate under the integral. For smooth factors, a locally integrable dominating derivative is sufficient. Hard cuts are treated weakly using boundary integrals; singular or coincident boundaries need a regularized selection law rather than formal products of distributions. A stochastic pipeline is represented by its inclusion probabilities before differentiation, as in Appendix~\ref{app:selection_response}.

\subsection{Complete first-order derivative}
\label{subsec:first_order_pair_decomposition}

At zero boost let
\begin{align}
\Lambda_{ab}=\rho_{2,ab}'\mathcal W_a\mathcal W_b\mathcal C_{ab},\qquad
\dot{\mathcal W}_{a,1}=\mathcal B_{\bm y_1}\mathcal W_a,
\end{align}
with the analogous definition for endpoint 2. Suppressing arguments common to the integrands, direct differentiation gives
\begin{align}
\delta_{\bm\beta}P_{ab}[K]
={}&\delta_{\bm\beta}P_{ab}^{\rm jac}[K]
+\delta_{\bm\beta}P_{ab}^{\rm ab}[K]
+\delta_{\bm\beta}P_{ab}^{\rm sel}[K]
\nonumber\\*
&+\delta_{\bm\beta}P_{ab}^{\rm pair}[K]
+\delta_{\bm\beta}P_{ab}^{K}[K],
\label{eq:pair_response_decomposition}
\end{align}
where all five terms are specified by
\begin{align}
&\delta_{\bm\beta}P_{ab}^{\rm jac}[K]
=2\int K_{ab}\Lambda_{ab}(\epsilon_1+\epsilon_2)\,\pd\Gamma_{12},
\label{eq:pair_jacobian_response}\\
&\delta_{\bm\beta}P_{ab}^{\rm ab}[K]
=-\int K_{ab}\mathcal W_a\mathcal W_b\mathcal C_{ab}\nonumber\\*
&\qquad\times\left(\bm\beta_{\perp1}\cdot\nabla_{S,1}+\bm\beta_{\perp2}\cdot\nabla_{S,2}\right)\rho_{2,ab}'\,\pd\Gamma_{12},
\label{eq:pair_aberration_response}\\
&\delta_{\bm\beta}P_{ab}^{\rm sel}[K]
=\int K_{ab}\rho_{2,ab}'\mathcal C_{ab}\nonumber\\*
&\qquad\times\left(\epsilon_1\dot{\mathcal W}_{a,1}\mathcal W_b+\epsilon_2\mathcal W_a\dot{\mathcal W}_{b,2}\right)\,\pd\Gamma_{12},
\label{eq:pair_endpoint_selection_response}\\
&\delta_{\bm\beta}P_{ab}^{\rm pair}[K]
=\int K_{ab}\rho_{2,ab}'\mathcal W_a\mathcal W_b\,\delta_{\bm\beta}\mathcal C_{ab}\,\pd\Gamma_{12},
\label{eq:pair_completeness_response}\\
&\delta_{\bm\beta}P_{ab}^{K}[K]
=\int (\delta_{\bm\beta}K_{ab})\Lambda_{ab}\,\pd\Gamma_{12}.
\label{eq:pair_kernel_response}
\end{align}
The derivatives of $\mathcal C_{ab}$ and $K_{ab}$ include their mark dependence and any explicit boost dependence. If a pair factor depends only on the two transformed endpoint marks, then
\begin{align}
\delta_{\bm\beta}\mathcal C_{ab}
&=\epsilon_1\dot{\mathcal C}_{ab}^{[1]}+\epsilon_2\dot{\mathcal C}_{ab}^{[2]},\qquad
\dot{\mathcal C}_{ab}^{[i]}=\mathcal B_{\bm y_i}\mathcal C_{ab}.
\label{eq:pair_completeness_endpoint_decomposition}
\end{align}
An irreducible pair-level response is retained as an additional $\delta_{\bm\beta}\mathcal C_{ab}$ in Eq.~\eqref{eq:pair_completeness_response}; it need not have the form in Eq.~\eqref{eq:pair_completeness_endpoint_decomposition}. Stochastic catalog decisions with correlated measurement errors require their joint inclusion law, rather than a product of marginal probabilities.

The disconnected functional is differentiated using the two selected one-point intensities and the same kernel:
\begin{align}
&\delta_{\bm\beta}M_{ab}[K]
=\int \left[(\delta_{\bm\beta}K)\lambda_a\lambda_b
+K(\delta_{\bm\beta}\lambda_a)\lambda_b\right.\nonumber\\*
&\hspace{5em}\left.+K\lambda_a(\delta_{\bm\beta}\lambda_b)\right]\,\pd Z_1\pd Z_2,
\label{eq:disconnected_general_derivative}\\
&\delta_{\bm\beta}C_{ab}[K]=\delta_{\bm\beta}P_{ab}[K]-\delta_{\bm\beta}M_{ab}[K].
\label{eq:connected_pair_response}
\end{align}
The one-point densities and kernel in this equation must use the same coordinate representation. Their boost derivatives cannot be replaced by constant coefficients when the angular selection or mark weights are nonuniform. The derivative of the signed population ratio follows by the quotient rule, but, as shown in Section~\ref{sec:normalization_contact}, does not generically cancel local density modulation.

\subsection{Equivalent coordinates and the role of the angular Jacobian}
\label{subsec:equivalent_pair_coordinates}

A change to two rest-frame directions moves the angular transformation from $\rho_2'$ and the solid-angle Jacobians to the angular arguments of every observed factor. The identity behind this equivalence is
\begin{align}
&\nabla_S\cdot\bm\beta_\perp=-2\epsilon,\nonumber\\*
&\int H\left(2\epsilon\rho-\bm\beta_\perp\cdot\nabla_S\rho\right)\,\pd\Omega\nonumber\\*
&\qquad=\int\rho\,\bm\beta_\perp\cdot\nabla_S H\,\pd\Omega.
\label{eq:angular_transport_identity}
\end{align}
It follows from integration by parts on the sphere; fixed survey boundaries are included in $H$ as weak derivatives. The two-endpoint identity follows by summing the two generators. Thus it is the \emph{Jacobian plus density-aberration} contribution that is exchanged for angular transport of the observed factors. Density remapping alone is not equivalent to moving only the bin kernel.

For $u=\hat{\bm n}_1\cdot\hat{\bm n}_2$, the inverse map gives
\begin{align}
u'&=u-(1-u)(\epsilon_1+\epsilon_2)+\mathcal O(\beta^2),
\label{eq:aberrated_pair_cosine}\\
\theta'&=\theta+2\sin(\theta/2)\,\bm\beta\cdot\hat{\bm N}+\mathcal O(\beta^2).
\label{eq:aberrated_pair_angle}
\end{align}
In rest-frame integration variables, a fixed observed angular kernel instead acquires $(1-u)(\epsilon_1+\epsilon_2)\partial_u K$. These are alternative representations of the same angular transport, not two additive physical distortions. Mark-induced changes of weights or radial bins remain distinct and must still be differentiated. Appendix~\ref{app:pair_response} gives the endpoint derivation and the hard-bin boundary term.

\subsection{Normalized angular response}
\label{subsec:local_angular_correlation_response}

For the local result, assume an isotropic parent, angularly uniform objectwise selection, a rotationally invariant pair factor defined in observed geometry, and the endpoint response in Eq.~\eqref{eq:pair_completeness_endpoint_decomposition}. Write
\begin{align}
\mathcal N_{ab}^{(2)}(u)&=\int\rho_{2,ab}'(u,\bm y_1,\bm y_2)\mathcal W_a\mathcal W_b\mathcal C_{ab}\,\pd\bm y_1\pd\bm y_2,\\
1+w_{ab}(u)&=\frac{\mathcal N_{ab}^{(2)}(u)}{n_an_b},\qquad
n_a=\int f_a(\bm y)\mathcal W_a(\bm y)\,\pd\bm y.
\label{eq:local_angular_correlation_definition}
\end{align}
For a retained pair list, $w_{ab}$ means excess over this explicitly specified product reference; it is not automatically a source-map covariance. Define the one-point and pair-conditioned selection scores by
\begin{align}
\sigma_a&=\frac{\int f_a\dot{\mathcal W}_a\,\pd\bm y}{n_a}=\mathcal A_a-2,\\
\sigma_{a\mid ab}^{[1]}(u)&=\frac{\int\rho_{2,ab}'\dot{\mathcal W}_a\mathcal W_b\mathcal C_{ab}\,\pd\bm y_1\pd\bm y_2}{\mathcal N_{ab}^{(2)}(u)}.
\label{eq:first_pair_conditioned_selection_score}
\end{align}
The second endpoint has the corresponding expression with $\mathcal W_a\dot{\mathcal W}_b$. The pair-completeness and geometrical scores are
\begin{align}
\kappa_{ab}^{[i]}(u)&=\frac{\int\rho_{2,ab}'\mathcal W_a\mathcal W_b\dot{\mathcal C}_{ab}^{[i]}\,\pd\bm y_1\pd\bm y_2}{\mathcal N_{ab}^{(2)}(u)},\\
\mathcal G_{ab}(u)&=-\frac{1-u}{\mathcal N_{ab}^{(2)}(u)}\int(\partial_u\rho_{2,ab}')\mathcal W_a\mathcal W_b\mathcal C_{ab}\,\pd\bm y_1\pd\bm y_2.
\label{eq:pair_geometrical_score}
\end{align}
The derivative of $\mathcal C_{ab}$ with respect to observed $u$ is not part of $\mathcal G_{ab}$, since that argument is held fixed. If a different pair geometry is used, its transformation belongs in $\delta\mathcal C_{ab}$ or in the kernel.

Set
\begin{align}
\Delta\Sigma_{ab}^{[1]}&=\sigma_{a\mid ab}^{[1]}+\kappa_{ab}^{[1]}-\sigma_a,\\
\Delta\Sigma_{ab}^{[2]}&=\sigma_{b\mid ab}^{[2]}+\kappa_{ab}^{[2]}-\sigma_b.
\label{eq:pair_selection_residuals}
\end{align}
The two angular Jacobians cancel against the two \emph{local} one-point means, leaving
\begin{align}
\delta_{\bm\beta}w_{ab}
={}&(1+w_{ab})\left\{(\epsilon_1+\epsilon_2)\mathcal G_{ab}\right.\nonumber\\*
&\left.\qquad+\epsilon_1\Delta\Sigma_{ab}^{[1]}+\epsilon_2\Delta\Sigma_{ab}^{[2]}\right\}.
\label{eq:local_angular_response_master}
\end{align}
All scores are evaluated at $u=u_{12}$. An additional irreducible pair response contributes $(n_an_b)^{-1}\int\rho_2'\mathcal W_a\mathcal W_b\delta\mathcal C_{\rm irr}\,\pd\bm y_1\pd\bm y_2$. The score ratios require $n_a,n_b,\mathcal N^{(2)}>0$. At a zero of the pair measure, the well-defined starting point is $\delta w=\delta\mathcal N^{(2)}/(n_an_b)-(1+w)(\delta\ln n_a+\delta\ln n_b)$, without dividing by $\mathcal N^{(2)}$.

\subsection{Normalized-measure identity and mark covariance}
\label{subsec:mark_clustering_covariance}

For smooth objectwise selection define
\begin{align}
p_{a,\varepsilon}(\bm y)&=\frac{f_a(\bm y)\mathcal W_{a,\varepsilon}(\bm y)}{I_a(\varepsilon)},\qquad
s_a=\left.\partial_\varepsilon\ln\mathcal W_{a,\varepsilon}\right|_0,\\
\left.\partial_\varepsilon p_{a,\varepsilon}\right|_0
&=p_a(s_a-\langle s_a\rangle_a).
\label{eq:normalized_mark_score_identity}
\end{align}
Here $I_a=\int f_a\mathcal W_{a,\varepsilon}$, and the parent density is held fixed. For an integrable, possibly explicitly varying quantity $g_\varepsilon$,
\begin{align}
\left.\partial_\varepsilon\langle g_\varepsilon\rangle_{a,\varepsilon}\right|_0
=\langle\dot g\rangle_a+\cov_a(g,s_a).
\label{eq:score_expectation_derivative}
\end{align}
The equation follows by differentiating the normalizing integral as well as the numerator. The following application identifies the relevant catalog-dependent weighting.

Write $\rho_{2,ab}'=f_af_b\chi_{ab}$, where $\chi_{ab}=1+\xi_{ab}'$ is the marked pair-density ratio, and let $\langle\cdot\rangle_0$ average over $p_a(\bm y_1)p_b(\bm y_2)$. For $\mathcal C_{ab}=1$,
\begin{align}
1+w_{ab}(u)&=\langle\chi_{ab}(u)\rangle_0,\\
\Delta\Sigma_{ab}^{[1]}(u)&=\frac{\cov_0(s_a,\chi_{ab})}{1+w_{ab}(u)},\\
\Delta\Sigma_{ab}^{[2]}(u)&=\frac{\cov_0(s_b,\chi_{ab})}{1+w_{ab}(u)}.
\label{eq:selection_mark_covariance}
\end{align}
The constant part of $\chi$ drops out of the covariance. This expresses a joint mark--location dependence \citep{2026MNRAS.549g1015T}. Independence of the score and pair excess is sufficient, but not necessary, for cancellation: a vanishing covariance suffices at this order. For a general pair factor, a nonsingular numerator form is
\begin{align}
&[\delta_{\bm\beta} w_{ab}]_{\rm sel+pair}\nonumber\\*
&\quad=\sum_{i=1}^2\epsilon_i\left\{\cov_0(s_i,\mathcal C_{ab}\chi_{ab})
+\langle\dot{\mathcal C}_{ab}^{[i]}\chi_{ab}\rangle_0\right\}.
\label{eq:general_pair_covariance_numerator}
\end{align}
The symbols $s_1,s_2$ mean the two endpoint scores. Equation~\eqref{eq:general_pair_covariance_numerator} assumes the endpoint decomposition of $\delta\mathcal C$; an irreducible response is added separately. For a hard threshold, $\dot{\mathcal W}$ is a boundary distribution, not a pointwise random score $\delta_{\rm D}/\Theta$. One must use the unnormalized boundary numerators in Appendix~\ref{app:selection_response}, or a controlled smooth-selection limit.

\subsection{Effective bias and correspondence with previous calculations}
\label{subsec:prior_work_correspondence}

Let $b(\bm y)$ be an intrinsic reference bias held fixed under a selection perturbation. Applying Eq.~\eqref{eq:score_expectation_derivative} gives
\begin{align}
\bar b_\varepsilon&=\int p_\varepsilon(\bm y)b(\bm y)\,\pd\bm y,\\
\left.\partial_\varepsilon\bar b_\varepsilon\right|_0&=\cov(s,b),\qquad
\left.\partial_\varepsilon\ln\bar b_\varepsilon\right|_0=\frac{\cov(s,b)}{\bar b}.
\label{eq:effective_bias_derivative}
\end{align}
The logarithmic expression assumes $\bar b>0$. For $\bar b=0$ one retains the derivative of $\bar b$ rather than defining a fractional modulation. A bias inferred from transformed observables can have an explicit derivative, in which case $\langle\dot b\rangle$ must also be included.

For a sharp flux cut with arbitrary source response $q(S,\bm\eta)$, differentiation before taking the ratio gives
\begin{align}
\left.\partial_\varepsilon\bar b_\varepsilon\right|_0
={}&\frac{S_{\rm lim}}{n_0}\int f(S_{\rm lim},\bm\eta)q(S_{\rm lim},\bm\eta)\nonumber\\*
&\quad\times\left[b(S_{\rm lim},\bm\eta)-\bar b\right]\,\pd\bm\eta.
\label{eq:bias_boundary_response}
\end{align}
If $q$ is common to all sources, boosting the flux is equivalent to changing the threshold to $S_{\rm lim}e^{-q\varepsilon}$. Therefore
\begin{align}
\frac{\cov(s,b)}{\bar b}
=-q\frac{\partial\ln\bar b(S_{\rm lim})}{\partial\ln S_{\rm lim}}.
\label{eq:sharp_threshold_bias_correspondence}
\end{align}
Here the covariance is interpreted through the boundary limit for a hard cut. It is a representation of selection-dependent effective bias, not an independent extra effect to add to the same bias derivative.

For comparison, Eq.~(2.23) of \citet{2024JCAP...06..045L} reads, in their density-only linear-growth approximation,
\begin{align}
\frac{F_{\rm kin}-D_{\rm kin}}{\beta}
=-f+\frac{\partial\ln b}{\partial\ln(1+z)}
+\frac{2}{r\mathcal H}\frac{\partial\ln b}{\partial\ln S_*}.
\label{eq:lacasa_bias_response}
\end{align}
Here $f$ is the growth rate and $\mathcal H$ is the conformal Hubble rate. Their Eqs.~(3.6) and (4.9)--(4.10) give aberration--modulation inference and its degeneracy; Appendix~D.3 treats nonlinear cross-spectra and flux-cut finite differences.

The correspondence is made without identifying different observational paths. For a family $\delta_g=b(z,S_*)D(z)\delta_*$, let $\eta_z=\partial_\varepsilon\ln(1+z)$ and $\eta_S=\partial_\varepsilon\ln S_*$ specify a deformation of its reference arguments. The chain rule gives
\begin{align}
\partial_\varepsilon\ln[bD]
=\eta_z\left[-f+\partial_{\ln(1+z)}\ln b\right]
+\eta_S\partial_{\ln S_*}\ln b.
\label{eq:general_bias_path_correspondence}
\end{align}
The effective differential operator in Eq.~\eqref{eq:lacasa_bias_response} corresponds to $\eta_z=1$ and $\eta_S=2/(r\mathcal H)$. Our purely radiometric, single-threshold deformation with no redshift-coordinate change instead has $\eta_z=0$ and $\eta_S=-q$, giving Eq.~\eqref{eq:sharp_threshold_bias_correspondence}. The coefficients differ because the paths differ; they cannot be equated by matching the name of a flux threshold alone.

At fixed redshift in the convention of Eq.~\eqref{eq:lacasa_bias_response}, subtraction of the count dipole and division by the local boosted mean change the fluctuation modulation from $F_{\rm kin}/\beta$ to $(F_{\rm kin}-D_{\rm kin})/\beta$. Section~\ref{sec:normalization_contact} derives this normalization step in the catalog notation. A projected angular sample additionally requires its redshift selection and marked covariance to be integrated before normalization. Thus the present theory includes the single-cut bias response as a restricted case and extends the bookkeeping to general joint selection and pair prescriptions; it does not claim to derive redshift evolution or light-cone dynamics from angular aberration alone.

\subsection{Linear-bias mixture and a calibration counterexample}
\label{subsec:linear_bias_selection_model}

Assume independent objectwise selection, no additional pair factor, and a common underlying field with angular correlation $w_{\rm m}$. Selected classes have probabilities $p(c)$, scores $s(c)$, and fixed biases $b(c)$, with marked pair-density ratio
\begin{align}
\chi_{12}(\theta)=1+b(c_1)b(c_2)w_{\rm m}(\theta).
\label{eq:linear_bias_pair_excess}
\end{align}
This is a rank-one clustering model. It is either a large-scale perturbative approximation or a realizable Cox model with a bounded latent field and positive conditional intensities; the validation below uses the latter. Let $\bar b=\langle b\rangle>0$. Substituting Eq.~\eqref{eq:linear_bias_pair_excess} into Eq.~\eqref{eq:selection_mark_covariance} yields
\begin{align}
w(\theta)&=\bar b^2w_{\rm m}(\theta),\\
\Delta\Sigma(\theta)&=\frac{\bar b\,w_{\rm m}(\theta)\cov(s,b)}{1+\bar b^2w_{\rm m}(\theta)}.
\label{eq:linear_bias_selection_residual}
\end{align}
The locally normalized midpoint response is
\begin{align}
\mathcal R_{aa}^{\rm mid}(\theta)
={}&2\sin(\theta/2)\bar b^2\frac{\pd w_{\rm m}}{\pd\theta}\nonumber\\*
&+2\cos(\theta/2)\bar b\,w_{\rm m}\cov(s,b),
\label{eq:linear_bias_midpoint_response}\\
\mathcal M_{\rm sel}&\equiv\frac{\cov(s,b)}{\bar b}.
\label{eq:effective_selection_modulation}
\end{align}
The selection term is $2\cos(\theta/2)\mathcal M_{\rm sel}w$. For two classes,
\begin{align}
\cov(s,b)=p(1-p)(s_1-s_2)(b_1-b_2).
\label{eq:two_class_covariance}
\end{align}
With $0<p<1$, this covariance is zero exactly when a score difference or a bias difference vanishes. The observable response may also vanish at a zero of $w_{\rm m}$; these are different statements.

To demonstrate the insufficiency of a count-only calibration, take $p=1/2$, $(b_1,b_2)=(1,3)$, and compare $(s_1,s_2)=(0,2)$ with $(2,0)$. Both cases have the same selected density, $\langle s\rangle=1$, $\mathcal A=3$, $\bar b=2$, and $w=4w_{\rm m}$. Yet
\begin{align}
\mathcal M_{\rm sel}^{\rm I}=+\frac12,\qquad
\mathcal M_{\rm sel}^{\rm II}=-\frac12.
\label{eq:counterexample_opposite_modulations}
\end{align}
These scores are realized by the valid smooth probabilities $\mathcal W_{c,\varepsilon}=[1+\exp(-2s_c\varepsilon)]^{-1}$, for which $\mathcal W_{c,0}=1/2$ and $\partial_\varepsilon\ln\mathcal W_{c,\varepsilon}|_0=s_c$. They can represent class-dependent completeness slopes in a log-flux variable shifted by $\varepsilon$. The example fixes both the one-point calibration and the isotropic two-point correlation while changing their joint response; it is not a claim that bias dependence on selection was previously unknown.

\begin{figure*}[t]
\centering
\includegraphics[width=0.73\textwidth]{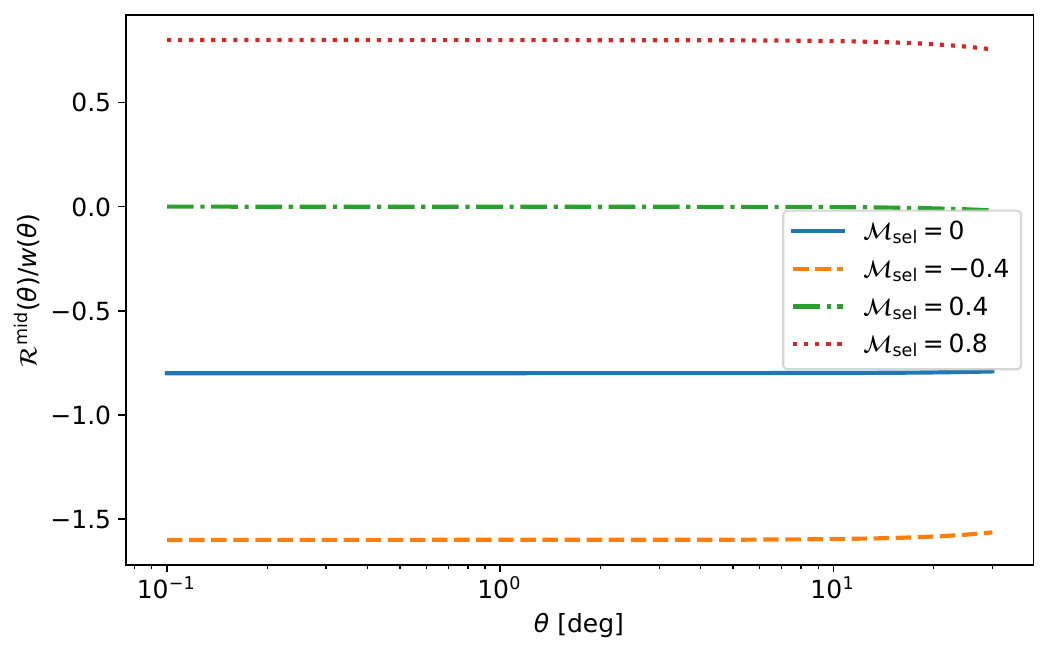}
\caption{Locally normalized midpoint response for a restricted-angle power law with $\gamma_w=0.8$. Distinct line styles identify the four selection-response coefficients. The small-angle cancellation occurs at $\mathcal M_{\rm sel}=\gamma_w/2$; finite-angle geometry leaves a small residual. This is a shape illustration, not a survey forecast or a global spherical covariance model.}
\label{fig:configuration_selection_coupling}
\end{figure*}

For a power law $w(\theta)\propto\theta^{-\gamma_w}$ on a restricted angular interval,
\begin{align}
\frac{\mathcal R^{\rm mid}}{w}
&=-2\gamma_w\frac{\sin(\theta/2)}{\theta}+2\mathcal M_{\rm sel}\cos(\theta/2),\\
\frac{\mathcal R_{\rm sel}^{\rm mid}}{\mathcal R_{\rm geom}^{\rm mid}}
&\simeq-\frac{2\mathcal M_{\rm sel}}{\gamma_w}\quad(\theta\ll1).
\label{eq:selection_to_geometry_ratio}
\end{align}
Figure~\ref{fig:configuration_selection_coupling} illustrates reinforcement, cancellation, and reversal. The power law is a local shape illustration, not an asserted covariance model on the entire sphere.

\subsection{Allowed midpoint and ordered-pair channels}
\label{subsec:auto_cross_response_channels}

For a same-tracer auto-correlation with exchange-symmetric selection, $\Delta\Sigma_{aa}^{[1]}=\Delta\Sigma_{aa}^{[2]}$. Since $\epsilon_1+\epsilon_2=2\cos(\theta/2)\bm\beta\cdot\hat{\bm N}$, Eq.~\eqref{eq:local_angular_response_master} gives
\begin{align}
\delta_{\bm\beta}w_{aa}&=\bm\beta\cdot\hat{\bm N}\,\mathcal R_{aa}^{\rm mid}(\theta),\\
\mathcal R_{aa}^{\rm mid}&=2\cos(\theta/2)(1+w_{aa})(\mathcal G_{aa}+\Delta\Sigma_{aa}).
\end{align}
For $\mathcal C=1$ and zero score--clustering covariance, this reduces to
\begin{align}
\mathcal R_{aa}^{\rm mid}(\theta)=2\sin(\theta/2)\frac{\pd w_{aa}}{\pd\theta}.
\label{eq:auto_midpoint_geometrical_limit}
\end{align}
Distinct tracers can have unequal endpoint residuals. Their antisymmetric combination multiplies $\epsilon_1-\epsilon_2=2\sin(\theta/2)\bm\beta\cdot\hat{\bm s}$ and reverses sign when the tracer labels are interchanged. Appendix~\ref{app:pair_response} gives both coefficients. No exchange-odd ordinary auto-pair signal is generated.

\section{Observable Normalization and Discrete-Map Contact Terms}
\label{sec:normalization_contact}

The response in Eq.~\eqref{eq:local_angular_response_master} is a property of a specific local normalization. This section constructs observables that measure it and distinguishes them from globally normalized pair projections and raw harmonic products. Unless stated otherwise, the map formulas refer to an actual simple source process, not a separately retained pair list.

\subsection{Local and global fluctuation fields}
\label{subsec:local_global_fields}

Let $\mathsf n_{a,\bm\beta}(\hat{\bm n})=\sum_{i\in X_a}\delta_{S^2}(\hat{\bm n},\hat{\bm n}_i)$ be the angular counting density and $n_{a,\bm\beta}=\mathbb E\mathsf n_{a,\bm\beta}$ its ensemble mean. Define the two centered generalized fields
\begin{align}
\Delta_{a,\bm\beta}^{\rm loc}
&=\frac{\mathsf n_{a,\bm\beta}-n_{a,\bm\beta}}{n_{a,\bm\beta}},\\
\Delta_{a,\bm\beta}^{\rm glob}
&=\frac{\mathsf n_{a,\bm\beta}-n_{a,\bm\beta}}{\bar n_{a,0}}.
\label{eq:local_global_map_definitions}
\end{align}
The second definition subtracts the deterministic mean, including its dipole, but divides the stochastic fluctuation by the unboosted global monopole. The two fields obey the exact relation $\Delta^{\rm glob}=(n_{\bm\beta}/\bar n_0)\Delta^{\rm loc}$. With uniform baseline selection,
\begin{align}
\Delta_{a,\bm\beta}^{\rm glob}
=\left[1+\mathcal A_a\epsilon(\hat{\bm n})\right]\Delta_{a,\bm\beta}^{\rm loc}+\mathcal O(\beta^2).
\label{eq:local_global_map_relation}
\end{align}
Subtracting a dipole is therefore not the same operation as local division by the boosted mean.

For the signal in the rank-one mixture, average the class intensities at fixed underlying field. Differentiation of $\bar b_\varepsilon$ and inverse aberration give
\begin{align}
\Delta_{a,\bm\beta}^{\rm loc,sig}
={}&\Delta_a^{\rm sig}-\nabla_S\epsilon\cdot\nabla_S\Delta_a^{\rm sig}\nonumber\\*
&+\mathcal M_{\rm sel}\epsilon\Delta_a^{\rm sig}+\mathcal O(\beta^2),\\
\Delta_{a,\bm\beta}^{\rm glob,sig}
={}&\Delta_a^{\rm sig}-\nabla_S\epsilon\cdot\nabla_S\Delta_a^{\rm sig}\nonumber\\*
&+(\mathcal A_a+\mathcal M_{\rm sel})\epsilon\Delta_a^{\rm sig}+\mathcal O(\beta^2).
\label{eq:normalization_specific_signal_maps}
\end{align}
These expressions describe the conditional intensity signal; a catalog with random membership is not obtained by multiplying each realization of shot noise by the same signal coefficient. Contact terms are derived separately below. In the redshift-resolved notation of Eq.~\eqref{eq:lacasa_bias_response}, the counterparts are $\mathcal A_z=D_{\rm kin}/\beta$, $\mathcal M_z^{\rm glob}=F_{\rm kin}/\beta$, and $\mathcal M_z^{\rm loc}=(F_{\rm kin}-D_{\rm kin})/\beta$, with the observational definition held fixed.

The ensemble mean in Eq.~\eqref{eq:local_global_map_definitions} is a specified calibration quantity, not the density of the particular realization being analyzed. If it is fitted from the same data, mean estimation, integral constraints, and correlations with the estimator must be included. One may instead forward-model a globally normalized map and retain its known one-point response. Local normalization is not a way to remove unknown velocity dependence without calibration.

\subsection{Why a scalar random-pair normalization is insufficient}
\label{subsec:boosted_random_normalization}

For the population Landy--Szalay ratio with $r_a=\lambda_{a,0}$ at the reference point, differentiation gives
\begin{align}
\delta_{\bm\beta}\Xi_{ab}^{\rm LS}[K\mid K_0]
={}&\frac{\delta_{\bm\beta}C_{ab}[K]}{M_{ab}[K_0]}\nonumber\\*
&-\Xi_{ab}[K\mid K_0]\frac{\delta_{\bm\beta}RR_{ab}[K_0]}{RR_{ab}[K_0]}.
\label{eq:LS_estimator_response_general}
\end{align}
All quantities on the right are evaluated at zero boost, and the derivative of $C$ includes any explicit kernel derivative. Fixed randoms give $\delta RR=0$, whereas response-matched randoms give $\delta RR=\delta M$. This quotient rule alone does not imply that the resulting observable measures $\delta w^{\rm loc}$.

For a same-tracer angular covariance write $c_{\bm\beta}(1,2)=n_{\bm\beta}(1)n_{\bm\beta}(2)w_{\bm\beta}(1,2)$. At uniform baseline density,
\begin{align}
\delta_{\bm\beta}c
=\bar n_0^2\left[\delta_{\bm\beta}w+\mathcal A(\epsilon_1+\epsilon_2)w_0\right].
\label{eq:connected_density_vs_local_response}
\end{align}
At fixed $0<\theta<\pi$, take a full-sky signed kernel $K=K_0\hat{\bm q}\cdot\hat{\bm N}$, where $K_0$ selects the separation and gives uniform pair-orientation weight. The isotropic target has $\Xi_0=0$. Since
\begin{align}
\langle(\hat{\bm q}\cdot\hat{\bm N})(\bm\beta\cdot\hat{\bm N})\rangle_{\rm orient}
=\frac{\hat{\bm q}\cdot\bm\beta}{3},
\end{align}
Eqs.~\eqref{eq:LS_estimator_response_general} and \eqref{eq:connected_density_vs_local_response} give
\begin{align}
\delta_{\bm\beta}\Xi_\theta
=\frac{\bm\beta\cdot\hat{\bm q}}{3}
\left[\mathcal R^{\rm mid}(\theta)+2\mathcal A\cos(\theta/2)w_0(\theta)\right].
\label{eq:integrated_midpoint_extra_modulation}
\end{align}
The scalar denominator term in Eq.~\eqref{eq:LS_estimator_response_general} is zero because it multiplies $\Xi_0$. Therefore the extra one-point term remains even for response-matched randoms. For a finite separation bin, Eq.~\eqref{eq:integrated_midpoint_extra_modulation} is replaced by its baseline pair-weighted bin average. Nonuniform orientation coverage requires the corresponding angular response matrix rather than $1/3$.

To project the locally normalized correlation, define an angular test function $\Psi$ and the inverse-intensity kernel
\begin{align}
K_{\bm\beta}^{\rm loc}(1,2)&=\frac{\Psi(1,2)}{n_{a,\bm\beta}(1)n_{b,\bm\beta}(2)},\\
\mathscr X_{\bm\beta}^{\rm loc}[\Psi\mid\Psi_0]
&=\frac{C_{\bm\beta}[K_{\bm\beta}^{\rm loc}]}{\int\Psi_0\,\pd\Omega_1\pd\Omega_2}\nonumber\\*
&=\frac{\int\Psi w_{ab,\bm\beta}\,\pd\Omega_1\pd\Omega_2}{\int\Psi_0\,\pd\Omega_1\pd\Omega_2}.
\label{eq:local_projection_kernel}
\end{align}
The angular marks have been marginalized here. The positive normalization is boost independent, and $\delta K^{\rm loc}/K^{\rm loc}=-(\delta\ln n_a+\delta\ln n_b)$ cancels the intensity factors before integration. This is the kernel counterpart of $\Delta^{\rm loc}$ and, unlike an unweighted ratio of pair counts, measures the local response. Division requires positive intensities on the analysis support; masked regions are excluded. Uncertain or data-dependent intensities remain part of the calibration problem.

\subsection{A reference matched to retained pairs}
\label{subsec:matched_pair_reference}

A pair-list selection introduces an additional reference choice. With $p_a,p_b$ the normalized selected one-point mark densities, define
\begin{align}
R_{ab}^{\mathcal C}(u)&=\langle\mathcal C_{ab}\rangle_0,\\
1+w_{ab}^{\mathcal C}(u)&=\frac{\langle\mathcal C_{ab}\chi_{ab}\rangle_0}{\langle\mathcal C_{ab}\rangle_0}
=\langle\chi_{ab}\rangle_{\mathcal C}.
\label{eq:pair_matched_reference}
\end{align}
The pair reference distribution is $p_ap_b\mathcal C/\langle\mathcal C\rangle_0$. This differs from the product-normalized $1+w_{ab}=\langle\mathcal C\chi\rangle_0$ in Eq.~\eqref{eq:local_angular_correlation_definition}. For an unclustered source population, $w_{ab}^{\mathcal C}=0$, while the product-normalized retained-pair statistic has $w_{ab}=\langle\mathcal C\rangle_0-1$. Both are legitimate operational targets, but they answer different questions.

For smooth positive $\mathcal C$ at fixed observed $u$, the normalized-measure identity gives
\begin{align}
\delta_{\bm\beta}w_{ab}^{\mathcal C}
={}&-(1-u)(\epsilon_1+\epsilon_2)\langle\partial_u\chi_{ab}\rangle_{\mathcal C}
\nonumber\\*
&+\cov_{\mathcal C}\left(\chi_{ab},\epsilon_1s_a+\epsilon_2s_b+\delta_{\bm\beta}\ln\mathcal C_{ab}\right).
\label{eq:pair_reference_covariance_response}
\end{align}
The first term differentiates the parent pair geometry, not the fixed observed argument of $\mathcal C$. Equation~\eqref{eq:pair_reference_covariance_response} includes an irreducible smooth pair response in $\delta\ln\mathcal C$. At zeros or hard boundaries, differentiate the numerator and denominator directly instead of forming a logarithmic score. A random catalog used for this target must represent the same joint pair reference, not only the two one-point intensities.

\subsection{Contact covariance and its boost response}
\label{subsec:contact_covariance}

For any simple angular point process, the ordinary and factorial second moments obey
\begin{align}
\cov[\mathsf n(1),\mathsf n(2)]
=c(1,2)+n(1)\delta_{S^2}(1,2).
\label{eq:point_process_contact_identity}
\end{align}
This identity does not require a Poisson process. A Poisson sampling interpretation identifies the second term with discreteness noise; non-Poisson correlations remain in $c$. For the two fields in Eq.~\eqref{eq:local_global_map_definitions}, the contact terms are
\begin{align}
\mathscr S_{\bm\beta}^{\rm loc}(1,2)&=\frac{\delta_{S^2}(1,2)}{n_{\bm\beta}(1)},\\
\mathscr S_{\bm\beta}^{\rm glob}(1,2)&=\frac{n_{\bm\beta}(1)}{\bar n_0^2}\delta_{S^2}(1,2).
\label{eq:normalization_contact_terms}
\end{align}
Set $N_0=1/\bar n_0$. Their first-order coefficients are respectively $-\mathcal A N_0\epsilon$ and $+\mathcal A N_0\epsilon$. For a boost along the $z$ axis, the harmonic contact response is
\begin{align}
\delta\langle a_{\ell m}a_{\ell+1,m}^*\rangle_{\rm contact}^{\rm loc}
&=-\mathcal A\beta N_0G_{\ell+1,m},\\
\delta\langle a_{\ell m}a_{\ell+1,m}^*\rangle_{\rm contact}^{\rm glob}
&=+\mathcal A\beta N_0G_{\ell+1,m},
\label{eq:contact_neighbor_response}
\end{align}
where $G_{\ell m}=[(\ell^2-m^2)/(4\ell^2-1)]^{1/2}$. These terms cannot be found by blindly replacing the signal spectrum by a total spectrum in a signal-only transformation law. Their sign is fixed by the normalization of the counting field.

For mark weights $\omega$, let $J_{2,\bm\beta}(\hat{\bm n})=\int\lambda_{\bm\beta}(\hat{\bm n},\bm y)\omega_{\bm\beta}^2\,\pd\bm y$ and let $d_{\bm\beta}$ be the map's deterministic normalizer. Then the contact amplitude is $J_{2,\bm\beta}/d_{\bm\beta}^2$, with logarithmic derivative $\delta\ln J_2-2\delta\ln d$. The squared-weight response is distinct from the mean weighted-count response. For overlapping tracers, replace $\omega^2/d^2$ by the product of the two weights and normalizers over objects common to both samples.

\subsection{Factorial harmonic estimators and covariance scope}
\label{subsec:factorial_harmonic_estimator}

Let $a_{\ell m}=\int\Delta^{\rm loc}Y_{\ell m}^*\,\pd\Omega$. The sample self-pair matrix is
\begin{align}
\widehat S_{\ell_1m_1,\ell_2m_2}
=\sum_{i\in X}\frac{Y_{\ell_1m_1}^*(\hat{\bm n}_i)Y_{\ell_2m_2}(\hat{\bm n}_i)}{n_{\bm\beta}(\hat{\bm n}_i)^2}.
\label{eq:sample_self_pair_matrix}
\end{align}
Its expectation is the harmonic transform of the first line of Eq.~\eqref{eq:normalization_contact_terms}. Thus
\begin{align}
\widehat V_{\ell_1m_1,\ell_2m_2}^{\rm fac}
=a_{\ell_1m_1}a_{\ell_2m_2}^*-
\widehat S_{\ell_1m_1,\ell_2m_2}
\label{eq:factorial_harmonic_product}
\end{align}
has the factorial connected correlation as its expectation. The deterministic mean terms in $a_{\ell m}$ are retained; they can be nonzero on a masked sky. For global normalization the self-pair weight is $1/\bar n_0^2$. For distinct tracers only identical objects shared by both catalogs are subtracted. These definitions are exact sample identities, not ensemble subtractions of an assumed isotropic noise level.

For a separately retained pair list, Eq.~\eqref{eq:factorial_harmonic_product} is replaced by the corresponding explicit distinct-pair sum with its reference terms. A general nonfactorizable pair weight cannot in general be reconstructed from a product of two map harmonics. Its expectation is still described by the pair-functional theory, but its covariance also depends on joint retention of pairs sharing objects and on retention of disjoint pairs.

Self-pair subtraction changes the mean but does not eliminate sampling variance. The exact covariance of factorial pair estimators contains up to fourth factorial moments and, for a retained pair list, joint pair-retention probabilities. Appendix~\ref{app:factorial_covariance} displays these terms. The simple Gaussian variance used later is exact for the stated full-sky Gaussian signal plus additive-noise model and is only an approximation for a discrete catalog after factorial subtraction. A realistic catalog application must calibrate its full covariance rather than equating these cases.

\section{Harmonic Template and Velocity-Estimator Bias}
\label{sec:harmonic_coupling}

The harmonic signal must refer to the same normalization, self-pair convention, and pair selection as the configuration-space target. We first transform a local factorial correlation, or the corresponding explicit pair projection. The factorial connected kernel need not itself be positive semidefinite, even for an actual source process; only its sum with the contact term is necessarily an ordinary covariance. A signal-covariance interpretation requires an additional continuous or conditional-intensity model, as in the rank-one Cox example. A separately retained pair target requires its explicit pair projection. We then specialize to the locally normalized signal field and distinguish raw-count contact noise from boost-invariant additive noise.

When such a signal decomposition is available, let $a_{\ell m}^a=\int\Delta_aY_{\ell m}^*\,\pd\Omega$ and write its isotropic \emph{signal} spectrum as
\begin{align}
\langle a_{\ell_1m_1}^aa_{\ell_2m_2}^{b*}\rangle_{0,\rm sig}
&=C_{\ell_1}^{ab}\delta_{\ell_1\ell_2}\delta_{m_1m_2},\\
w_{ab}(u)&=\sum_{j=0}^{\infty}\frac{2j+1}{4\pi}C_j^{ab}P_j(u).
\label{eq:signal_harmonic_definitions}
\end{align}
Self-pair contact terms are excluded from $C_j^{ab}$. Expand the boost field as
\begin{align}
\epsilon(\hat{\bm n})=\sum_{M=-1}^{1}\epsilon_{1M}Y_{1M}(\hat{\bm n}),\qquad
\epsilon_{1M}=\frac{4\pi}{3}\beta Y_{1M}^*(\hat{\bm\beta}).
\label{eq:velocity_harmonic_coefficients}
\end{align}
The coefficients satisfy $\epsilon_{1,-M}=(-1)^M\epsilon_{1M}^*$. The formulas refer to the finite multipole range actually analyzed; any monopole or dipole removal must be included in that range and in the survey operator.

\subsection{General pair response and BipoSH convention}
\label{subsec:generic_harmonic_pair_response}

Define
\begin{align}
\Delta\Sigma_{ab}^{(\pm)}&=\tfrac12(\Delta\Sigma_{ab}^{[1]}\pm\Delta\Sigma_{ab}^{[2]}),
\label{eq:response_residual_plus_minus}\\
\mathcal F_{ab}^{(+)}&=(1+w_{ab})(\mathcal G_{ab}+\Delta\Sigma_{ab}^{(+)}),\\
\mathcal F_{ab}^{(-)}&=(1+w_{ab})\Delta\Sigma_{ab}^{(-)}.
\end{align}
Then
\begin{align}
\delta_{\bm\beta}w_{ab}
=(\epsilon_1+\epsilon_2)\mathcal F_{ab}^{(+)}(u)
+(\epsilon_1-\epsilon_2)\mathcal F_{ab}^{(-)}(u).
\label{eq:harmonic_midpoint_separation_functions}
\end{align}
For a symmetric same-tracer target, $\mathcal F_{aa}^{(-)}=0$. Additional pair-response components, if present, must be transformed along with this local result. Expand
\begin{align}
\mathcal F_{ab}^{(\pm)}(u)
&=\sum_{j=0}^{\infty}\frac{2j+1}{4\pi}F_{j,ab}^{(\pm)}P_j(u),
\label{eq:response_function_legendre_expansion}\\
F_{j,ab}^{(\pm)}&=2\pi\int_{-1}^{1}\mathcal F_{ab}^{(\pm)}(u)P_j(u)\,\pd u.
\end{align}
The addition theorem and the integral of three spherical harmonics give
\begin{align}
&h_{\ell_1 1\ell_2}=\left[\frac{3(2\ell_1+1)(2\ell_2+1)}{4\pi}\right]^{1/2}
\begin{pmatrix}\ell_1&1&\ell_2\\0&0&0\end{pmatrix},
\label{eq:gaunt_prefactor_definition}\\
&\delta_{\bm\beta}\langle a_{\ell_1m_1}^aa_{\ell_2m_2}^{b*}\rangle_{\rm sig}
=\sum_M\epsilon_{1M}(-1)^{m_1}h_{\ell_1 1\ell_2}\nonumber\\*
&\qquad\times\begin{pmatrix}\ell_1&1&\ell_2\\-m_1&M&m_2\end{pmatrix}\mathcal Q_{\ell_1\ell_2}^{ab},
\label{eq:general_neighboring_covariance}\\
&\mathcal Q_{\ell_1\ell_2}^{ab}
=F_{\ell_1,ab}^{(+)}+F_{\ell_2,ab}^{(+)}+F_{\ell_2,ab}^{(-)}-F_{\ell_1,ab}^{(-)}.
\label{eq:general_harmonic_response_kernel}
\end{align}
The term proportional to $\epsilon_1$ selects the Legendre index $\ell_2$, and the term proportional to $\epsilon_2$ selects $\ell_1$. This fixes the relative signs in $\mathcal Q$. The triangle condition and the even parity of the Gaunt integral require
\begin{align}
\ell_2=\ell_1\pm1.
\label{eq:neighboring_multipole_selection}
\end{align}
This is an $L=1$ statistical anisotropy of the pair, not an exchange-odd ordinary separation multipole. The same linear transform applies to a retained-pair expectation, but then the left side is the corresponding harmonic pair projection rather than a covariance of raw maps.

For a chosen harmonic quadratic matrix $\widehat V$, define
\begin{align}
\widehat A_{\ell_1\ell_2}^{LM,ab}
={}&\sum_{m_1m_2}(-1)^{m_2}\widehat V_{\ell_1m_1,\ell_2m_2}^{ab}\nonumber\\*
&\quad\times\left\langle\ell_1m_1\,\ell_2,-m_2\middle|LM\right\rangle.
\label{eq:sample_BipoSH_estimator}
\end{align}
Use $\widehat V=aa^*$ for a continuous map, Eq.~\eqref{eq:factorial_harmonic_product} for a factorial source-catalog statistic, or an explicit weighted distinct-pair sum for a retained pair list. When its mean is the signal target in Eq.~\eqref{eq:general_neighboring_covariance}, orthogonality of the Clebsch--Gordan coefficients yields
\begin{align}
\langle\widehat A_{\ell_1\ell_2}^{1M,ab}\rangle
=(-1)^{\ell_1-\ell_2}\frac{h_{\ell_1 1\ell_2}}{\sqrt3}\mathcal Q_{\ell_1\ell_2}^{ab}\epsilon_{1M}
+\mathcal O(\beta^2).
\label{eq:biposh_template_definition}
\end{align}
For raw discrete-map products, the BipoSH contact mean must be added. Appendix~\ref{app:harmonic_details} provides the normalization and local-map derivation.

\subsection{Response fields beyond a constant modulation coefficient}
\label{subsec:response_field_cross_spectrum}

The selection response can be a field rather than a scalar times the selected fluctuation. For independently selected parent mark classes with contrast $\Delta(\hat{\bm n},\bm y)$, define the conditional-intensity signal fields
\begin{align}
\Delta_a^{\rm sig}(\hat{\bm n})&=\int p_a(\bm y)\Delta(\hat{\bm n},\bm y)\,\pd\bm y,\\
\mathcal U_a(\hat{\bm n})&=\int p_a(\bm y)[s_a(\bm y)-\sigma_a]\Delta(\hat{\bm n},\bm y)\,\pd\bm y.
\label{eq:selection_response_field}
\end{align}
Differentiating the normalized mark mixture gives
\begin{align}
\Delta_{a,\bm\beta}^{\rm loc,sig}
=\Delta_a^{\rm sig}-\nabla_S\epsilon\cdot\nabla_S\Delta_a^{\rm sig}
+\epsilon\mathcal U_a+\mathcal O(\beta^2).
\label{eq:general_response_field_map}
\end{align}
The cross-covariance $\langle\mathcal U_a(1)\Delta_b^{\rm sig}(2)\rangle$ is precisely the endpoint-1 covariance numerator in Eq.~\eqref{eq:selection_mark_covariance}. There is no need to assume linear bias, provided the required moments exist. Explicit intrinsic evolution or additional transformed-field terms, when part of the observable, add to $\mathcal U$; they are not contained in a pure selection score by definition.

For a real scalar auto-map, let $S_\ell=C_\ell^{\Delta,\mathcal U}$. A boost along $\hat{\bm z}$ gives
\begin{align}
&\delta\langle a_{\ell m}a_{\ell+1,m}^*\rangle_{\rm sig}\nonumber\\*
&\quad=\beta G_{\ell+1,m}\left[\ell C_\ell-(\ell+2)C_{\ell+1}+S_\ell+S_{\ell+1}\right].
\label{eq:general_cross_spectral_response}
\end{align}
This has the response cross-spectrum structure of Appendix~D.3 of \citet{2024JCAP...06..045L}. Here the response field is defined by general mark selection and the chosen normalization. In the rank-one model, $\mathcal U=\mathcal M_{\rm sel}\Delta$, hence $S_\ell=\mathcal M_{\rm sel}C_\ell$. For global normalization, replace $\mathcal U$ by $\mathcal U+\mathcal A\Delta$. A nonfactorizable pair-list prescription need not admit an objectwise response field; its response is retained through $\mathcal F^{(\pm)}$ or the full kernelized integral.

\subsection{Geometry and modulation templates}
\label{subsec:harmonic_selection_template}

When the signal does admit a constant modulation $\mathcal M_a$, Eqs.~\eqref{eq:normalization_specific_signal_maps} and \eqref{eq:general_cross_spectral_response} reduce to
\begin{align}
&\langle\widetilde a_{\ell m}^a\widetilde a_{\ell+1,m}^{a*}\rangle_{\rm sig}\nonumber\\*
&\quad=\beta G_{\ell+1,m}\left\{\ell C_\ell^{aa}-(\ell+2)C_{\ell+1}^{aa}\right.\nonumber\\*
&\hspace{7em}\left.+\mathcal M_a[C_\ell^{aa}+C_{\ell+1}^{aa}]\right\}
+\mathcal O(\beta^2),
\label{eq:auto_neighboring_covariance_local_map}\\
&G_{\ell m}=\left[\frac{\ell^2-m^2}{4\ell^2-1}\right]^{1/2}.
\label{eq:G_lm_definition_compact}
\end{align}
For the local rank-one signal,
\begin{align}
\mathcal M_a=\mathcal M_{\rm sel}=\frac{\cov(s,b)}{\bar b},
\label{eq:linear_bias_harmonic_modulation}
\end{align}
whereas for the global signal $\mathcal M_a=\mathcal A_a+\mathcal M_{\rm sel}$. These are the established aberration and modulation shapes \citep{2024JCAP...06..045L}, recovered here with an explicit catalog convention. Define
\begin{align}
T_\ell^{\rm geom}&=\ell C_\ell-(\ell+2)C_{\ell+1},\\
T_\ell^{\rm sel}&=C_\ell+C_{\ell+1},
\label{eq:geometry_selection_harmonic_shapes}\\
T_\ell^{\rm true}&=T_\ell^{\rm geom}+\mathcal M_{\rm sel}T_\ell^{\rm sel}
\label{eq:true_neighboring_template}
\end{align}
for the locally normalized rank-one signal. For raw local counts, add $-\mathcal A N_0$ inside $T_\ell^{\rm true}$; for raw global counts use $\mathcal A+\mathcal M_{\rm sel}$ and add $+\mathcal A N_0$. Mark weighting changes these contact templates according to Section~\ref{subsec:contact_covariance}. Simply using $C_\ell+N_0$ everywhere in Eq.~\eqref{eq:geometry_selection_harmonic_shapes} is not the appropriate substitution for a generally selected catalog.

For $C_\ell\propto\ell^{-\nu}$, $\ell\gg1$, and $\nu\neq2$,
\begin{align}
T_\ell^{\rm geom}&\simeq(\nu-2)C_\ell,\qquad T_\ell^{\rm sel}\simeq2C_\ell,\nonumber\\*
\frac{T_\ell^{\rm sel}}{T_\ell^{\rm geom}}&\simeq\frac{2}{\nu-2}.
\label{eq:template_powerlaw_degeneracy}
\end{align}
The templates are nearly proportional, not automatically distinguishable by their origin. At $\nu=2$ the leading geometrical term vanishes and the ratio is not a valid approximation. In fact $C_\ell\propto[\ell(\ell+1)]^{-1}$ gives $T_\ell^{\rm geom}=0$ exactly for the multipoles on which this form holds.

A finite-band exact proportionality example follows from the recurrence
\begin{align}
\frac{C_{\ell+1}}{C_\ell}=\frac{\ell-k}{\ell+2+k}
\quad\Longrightarrow\quad T_\ell^{\rm geom}=kT_\ell^{\rm sel}.
\label{eq:exact_template_degeneracy}
\end{align}
Positive spectra $C_\ell\propto\Gamma(\ell-k)/\Gamma(\ell+2+k)$ realize this relation on any analyzed band where the arguments are positive. The band may be embedded in a finite-resolution covariance. A broad multipole interval alone therefore does not guarantee identifiability.

\subsection{Quadratic inference and a misspecified template}
\label{subsec:geometry_only_estimator_bias}

For neighboring multipoles define
\begin{align}
\mathcal H_\ell&=-h_{\ell\,1\,\ell+1}/\sqrt3,\qquad
|\mathcal H_\ell|^2=(\ell+1)/(4\pi),\\
\mathcal T_\ell^{\rm geom}&=\mathcal H_\ell T_\ell^{\rm geom},\qquad
\mathcal T_\ell^{\rm sel}=\mathcal H_\ell T_\ell^{\rm sel},\\
\mathcal T_\ell^{\rm true}&=\mathcal T_\ell^{\rm geom}+\mathcal M_{\rm sel}\mathcal T_\ell^{\rm sel}.
\label{eq:all_biposh_templates}
\end{align}
We now specify the model underlying the simple weights. Take a continuous mean-zero Gaussian signal with the local response in Eq.~\eqref{eq:true_neighboring_template}, plus independent, isotropic, boost-invariant additive noise. Let
\begin{align}
C_\ell^{\rm tot}=C_\ell+N_\ell.
\label{eq:total_spectrum_definition}
\end{align}
The noise contributes no off-diagonal mean at first order. At the full-sky Gaussian null, Wick contraction and Clebsch--Gordan orthogonality give
\begin{align}
\operatorname{Var}(\widehat A_{\ell,\ell+1}^{1M})
=C_\ell^{\rm tot}C_{\ell+1}^{\rm tot}\equiv V_\ell,
\label{eq:BipoSH_estimator_variance}
\end{align}
where variance of a complex coefficient means its mean squared modulus. Different ordered neighboring blocks $\ell,\ell+1$ are uncorrelated in this null model. Reality relates the $M$ and $-M$ components; they must not be counted as independent complex data. For a factorial catalog statistic, Eq.~\eqref{eq:BipoSH_estimator_variance} is only a Gaussian approximation to its actual covariance, as discussed in Appendix~\ref{app:factorial_covariance}.

With fixed calibrated templates, positive $V_\ell$, and nonzero total template norm, inverse-variance weighting gives
\begin{align}
\widehat\epsilon_{1M}
&=\mathcal N_M\sum_{\ell=\ell_{\min}}^{\ell_{\max}-1}
\frac{\mathcal T_\ell^{\rm true*}\widehat A_{\ell,\ell+1}^{1M}}{V_\ell},
\label{eq:velocity_harmonic_estimator}\\
\mathcal N_M^{-1}
&=\sum_{\ell=\ell_{\min}}^{\ell_{\max}-1}\frac{|\mathcal T_\ell^{\rm true}|^2}{V_\ell}.
\label{eq:velocity_estimator_normalization}
\end{align}
It is unbiased through first order for this known template, with null variance $\mathcal N_M$. This is the minimum-variance linear combination of the stated quadratic coefficients under the stated covariance, not a claim of unconditional optimality for arbitrary catalogs. Calibration uncertainty in $C_\ell$, the selection response, or the noise mean requires nuisance parameters or independent validation.

Define the template inner product $\langle X,Y\rangle_V=\sum_\ell X_\ell^*Y_\ell/V_\ell$ and set $G=\mathcal T^{\rm geom}$, $S=\mathcal T^{\rm sel}$. For $\langle G,G\rangle_V>0$, using only $G$ gives
\begin{align}
\mathbb E\widehat\epsilon_{1M}^{\rm geom}
&=\left[1+B\mathcal M_{\rm sel}\right]\epsilon_{1M}+\mathcal O(\beta^2),\\
B&=\frac{\langle G,S\rangle_V}{\langle G,G\rangle_V}.
\label{eq:geometry_only_estimator_bias}
\end{align}
This is the standard template-mismatch projection, also discussed in Appendix~D.2 of \citet{2024JCAP...06..045L}, applied here to the specified selection response. It is valid without dividing by a velocity component that might be zero. An unmodeled contact or instrumental mean template contributes an additional projection of that template onto $G$.

\begin{figure*}[t]
\centering
\includegraphics[width=0.73\textwidth]{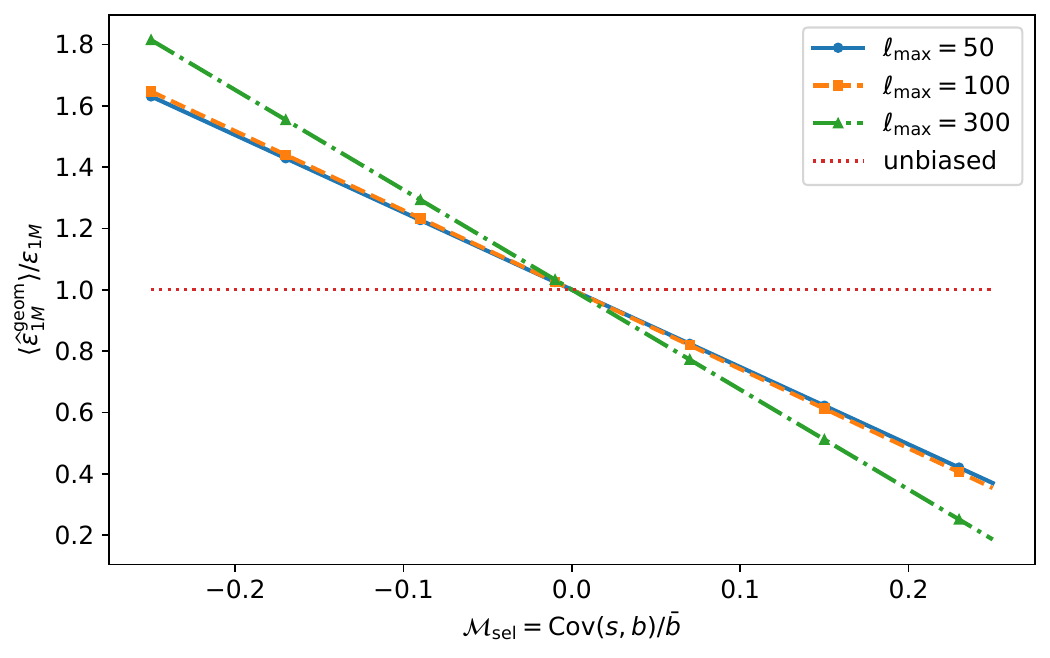}
\caption{Geometry-only velocity inference in the continuous Gaussian signal plus boost-invariant additive-noise model of Eq.~\eqref{eq:toy_harmonic_spectrum}. The same multiplicative factor applies to each nonzero velocity component under isotropic full-sky weighting. The plotted ratios follow Eq.~\eqref{eq:geometry_only_estimator_bias}; they are not forecasts for raw source-count maps, whose contact response must also be modeled.}
\label{fig:geometry_only_estimator_bias}
\end{figure*}

For illustration take
\begin{align}
C_\ell&\propto(\ell+\tfrac12)^{-1.2}\exp[-(\ell/600)^2],\nonumber\\*
N_\ell&=0.1C_{100},\qquad \ell_{\min}=2.
\label{eq:toy_harmonic_spectrum}
\end{align}
The amplitude cancels from $B$ when noise is scaled as stated. Figure~\ref{fig:geometry_only_estimator_bias} uses $\ell_{\max}=50,100,300$, with sums through $\ell_{\max}-1$. Its white noise is the boost-invariant additive noise just defined, not the contact noise of an uncorrected Poisson catalog. A factorial catalog can have the same mean template after self-pair subtraction, but its sampling covariance must be checked independently. The figure illustrates a possible inference bias rather than a survey forecast.

\subsection{Joint amplitudes and rank of the response}
\label{subsec:identifiability_rank}

A fit does not remove a degeneracy merely by introducing another parameter. For a fixed dipole direction, use two real amplitudes $x=\epsilon_{10}$ and $y=\mathcal M_{\rm sel}\epsilon_{10}$, so that the mean data vector is $Gx+Sy$. With the null covariance held fixed, the normal equations are
\begin{align}
\bm F\begin{pmatrix}\widehat x\\\widehat y\end{pmatrix}
&=\begin{pmatrix}\langle G,\widehat A\rangle_V\\\langle S,\widehat A\rangle_V\end{pmatrix},\\
\bm F&=\begin{pmatrix}\langle G,G\rangle_V&\langle G,S\rangle_V\\\langle S,G\rangle_V&\langle S,S\rangle_V\end{pmatrix}.
\label{eq:joint_amplitude_fisher}
\end{align}
This reproduces the two-template inference structure of \citet{2024JCAP...06..045L}. For real templates with nonzero individual norms define
\begin{align}
r_{GS}&=\frac{\langle G,S\rangle_V}{\sqrt{\langle G,G\rangle_V\langle S,S\rangle_V}},\\
\det\bm F&=\langle G,G\rangle_V\langle S,S\rangle_V(1-r_{GS}^2),\\
\frac{\sigma_{x,\rm joint}}{\sigma_{x,\,y\ fixed}}
&=\frac{1}{\sqrt{1-r_{GS}^2}}.
\label{eq:joint_estimator_degeneracy}
\end{align}
The denominator in the last line corresponds to fixing the independent nuisance amplitude $y$, not to fixing $\mathcal M_{\rm sel}$ along $y=\mathcal M_{\rm sel}x$. The latter uses the combined template and has a different conditional error. For an unknown direction one can fit independent dipole vectors $x_M,y_M$ as a conservative enlarged model, or enforce the shared-scalar relation $y_M=\mathcal M_{\rm sel}x_M$. If a template vanishes, the corresponding amplitude is unconstrained and the correlation coefficient is not defined. At zero velocity, $\mathcal M_{\rm sel}$ itself is not identifiable from this signal.

\begin{figure*}[t]
\centering
\includegraphics[width=0.73\textwidth]{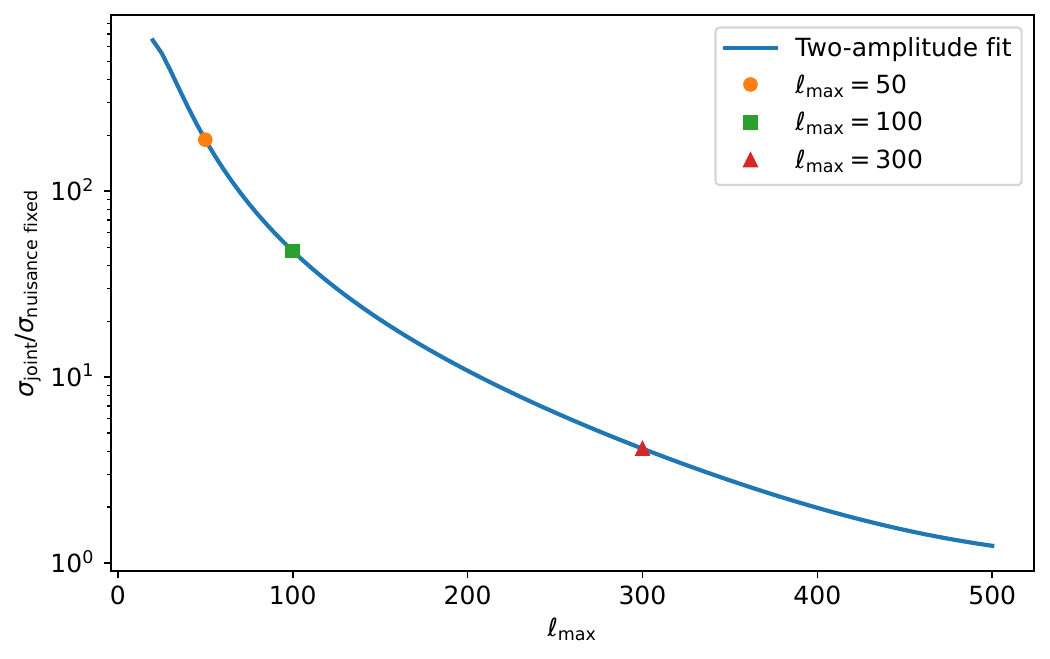}
\caption{Uncertainty increase when the independent selection amplitude is fitted jointly with the geometrical amplitude, rather than held fixed. The spectra, noise, and multipole sums are those of Fig.~\ref{fig:geometry_only_estimator_bias}. The error ratio is $(1-r_{GS}^2)^{-1/2}$, not a detection significance. For nearly scale-free bands the joint inference is poorly conditioned despite a large geometry-only bias.}
\label{fig:identifiability_cost}
\end{figure*}

Table~\ref{tab:template_degeneracy} quantifies the toy-model degeneracy. Large misspecification bias and poor internal separation are two aspects of the same template overlap. Figure~\ref{fig:identifiability_cost} shows the uncertainty cost as the analyzed band is extended. The gain comes from curvature in the chosen spectrum, not from the number of multipoles alone.
\begin{table}[t]
\centering
\caption{Template overlap in Eq.~\eqref{eq:toy_harmonic_spectrum}. The last column is the error ratio defined in Eq.~\eqref{eq:joint_estimator_degeneracy}.}
\label{tab:template_degeneracy}
\begin{tabular}{rrrr}
\toprule
$\ell_{\max}$ & $B$ & $r_{GS}$ & Error ratio\\
\midrule
50 & $-2.52139$ & $-0.999986$ & $189.33$\\
100 & $-2.58644$ & $-0.999780$ & $47.70$\\
300 & $-3.25715$ & $-0.970183$ & $4.126$\\
\bottomrule
\end{tabular}
\end{table}

\subsection{Survey operators and irreducible degeneracy}
\label{subsec:survey_window_response}

A fixed linear mask, beam, pixelization, or deterministic map weight can be represented by an operator $\bm W$. Its covariance response is
\begin{align}
\bm C_{\rm obs}&=\bm W\bm C_{\rm sky}\bm W^\dagger+\bm N,\\
\delta\bm C_{\rm obs}&=\bm W(\delta\bm C_{\rm sky})\bm W^\dagger+\delta\bm N.
\label{eq:windowed_response}
\end{align}
If $\bm W$ also changes with the modeled perturbation, its two product-rule derivatives must be included. Spatially varying depth can change the selected mark mixture, not only multiply a pre-existing map; that effect is calculated from the direction-dependent pair functional and cannot in general be compressed into one fixed $\bm W$.

For two covariance templates $\bm S=k\bm G$, a common fixed operator preserves proportionality: $\bm W\bm S\bm W^\dagger=k\bm W\bm G\bm W^\dagger$. Equivalently, no reweighting of identical data responses can increase the rank of a deficient design matrix. Masks and weights can alter a \emph{near} degeneracy by redistributing available information, but cannot break an exact one. Independent tracers, redshift information, additional response shapes, or an external calibration are required to add a missing direction in parameter space.

Practical inference must subtract the baseline anisotropic mean field, propagate all templates through the same catalog and map-making pipeline, and use the full covariance. The window does not invalidate the exchange theorem. It changes the allowed even-sector signal and its covariance; it does not turn an exchange-odd identical-endpoint statistic into an admissible physical dipole.

\section{Independent Forward-Model Validation}
\label{sec:validation}

The figures based on analytic templates illustrate their consequences, but are not by themselves independent tests of the response derivation. We therefore compare the linear formulas with centered differences of complete finite-boost expectation values. These calculations are deterministic quadratures of explicitly defined models. They do not simulate a survey, establish a detection significance, or validate a Gaussian covariance for sparse point catalogs.

\subsection{A realizable marked source population}
\label{subsec:bounded_cox_model}

Consider a Cox process with two intrinsic source classes. Conditional on a bounded random field $d_{\rm m}(\hat{\bm n})$, its class intensities are
\begin{align}
\lambda_c^{\rm parent}(\hat{\bm n})&=f_c[1+b_c d_{\rm m}(\hat{\bm n})],\\
d_{\rm m}(\hat{\bm n})&=a_0Z_0+a_1\hat{\bm e}\cdot\hat{\bm n},\qquad
w_{\rm m}(u)=a_0^2+\frac{a_1^2}{3}u.
\label{eq:bounded_cox_benchmark}
\end{align}
Here $Z_0=\pm1$ with equal probability and $\hat{\bm e}$ is an independent uniformly distributed unit vector. This ensemble is isotropic and has the rank-one marked pair-density ratio of Eq.~\eqref{eq:linear_bias_pair_excess}. Set $a_0=0.10$ and $a_1=0.20$, and use
\begin{align}
(f_1,f_2)&=(0.55,0.45),\qquad (x_1,x_2)=(-0.35,0.40),\\
(q_1,q_2)&=(1.1,2.0),\qquad (b_1,b_2)=(0.8,1.6).
\label{eq:benchmark_class_parameters}
\end{align}
The $f_c$ sum to a parent intensity of one per steradian. Since $|d_{\rm m}|\leq0.30$, every conditional intensity factor is at least $1-1.6(0.30)=0.52$. This avoids the negative-intensity problem of an unrestricted linear Gaussian Cox construction. The arbitrary overall intensity is irrelevant to these expectation-level derivative tests.

The mark $x_c$ is log flux in a reference unit and transforms as $x_{c,\delta}=x_c+q_c\ln\delta$, corresponding to a power-law radiometric response. Take the smooth object inclusion probability
\begin{align}
\mathcal W(x)=[1+\exp(0.2-1.4x)]^{-1}.
\label{eq:benchmark_object_selection}
\end{align}
The derivative is $\dot{\mathcal W}_c=1.4q_c\mathcal W_c(1-\mathcal W_c)$. All object probabilities remain between zero and one under the finite boosts used below.

\subsection{Nonfactorizable retention and two pair references}
\label{subsec:nonfactorizable_validation}

After object inclusion, retain each unordered pair with probability
\begin{align}
&\mathcal C_{\bm\beta}(u,c,d)\nonumber\\*
&\quad=\operatorname{logit}^{-1}\!\left[0.1+0.3u+0.7(x_{c,\delta_1}+x_{d,\delta_2})\right].
\label{eq:benchmark_pair_retention}
\end{align}
Here $\operatorname{logit}^{-1}(v)=(1+e^{-v})^{-1}$. This is a symmetric, nonfactorizable pair-list operation. The same retention variable is used for both orderings of a pair. Objects themselves are not removed by this step, so $\mathcal C$ changes the pair measure but not the one-point means. It is not asserted to be a factorial density of a new source process. At fixed observed $u$,
\begin{align}
\delta\mathcal C
=0.7\mathcal C(1-\mathcal C)(q_c\epsilon_1+q_d\epsilon_2).
\label{eq:benchmark_retention_derivative}
\end{align}
We test both the product-reference target in Eq.~\eqref{eq:local_angular_correlation_definition} and the retained-pair reference in Eq.~\eqref{eq:pair_matched_reference}.

For the local tests use
\begin{align}
\hat{\bm n}_1&=(1,2,3)/\sqrt{14},\qquad
\hat{\bm n}_2=(-2,1,4)/\sqrt{21},\\
\hat{\bm v}&=(0.3,-0.4,\sqrt{0.75}),\qquad \bm\beta=h\hat{\bm v}.
\label{eq:benchmark_geometry}
\end{align}
At each sign of $h$, we compute the exact Doppler factors, inverse-aberrated directions, boosted log fluxes, inclusion probabilities, pair-retention probabilities, and the appropriate denominator. For a unit vector $\hat{\bm v}$, $\mu=\hat{\bm v}\cdot\hat{\bm n}$, and signed speed $h$, the exact inverse direction is
\begin{align}
\hat{\bm n}'(h)
&=\frac{\hat{\bm n}+[(\gamma_h-1)\mu-\gamma_h h]\hat{\bm v}}{\gamma_h(1-h\mu)},\nonumber\\*
\gamma_h&=(1-h^2)^{-1/2}.
\label{eq:benchmark_exact_inverse_boost}
\end{align}
The numerical derivative $[F(h)-F(-h)]/(2h)$ is compared with the independently assembled first-order terms. The three local outcomes are in Table~\ref{tab:local_validation}. The large positive response of the product-normalized retained-pair statistic includes migration through the pair-retention rule; it is not evidence for a physical increase of matter clustering.
\begin{table*}[t]
\centering
\caption{Local expectation-level validation. Derivatives are with respect to signed speed $h$ along Eq.~\eqref{eq:benchmark_geometry}. The error column uses the fixed step $h=10^{-4}$, not a fitted or optimized step.}
\label{tab:local_validation}
\begin{tabular}{lrrr}
\toprule
Target & $w_0$ & Analytic derivative & Relative difference\\
\midrule
Objectwise selection, $\mathcal C=1$ & $0.03130280094$ & $-0.006464351016$ & $9.54\times10^{-9}$\\
Retained pairs, product reference & $-0.3746920484$ & $0.2961388863$ & $2.67\times10^{-9}$\\
Retained pairs, matched pair reference & $0.03323676868$ & $-0.007100795649$ & $1.09\times10^{-8}$\\
\bottomrule
\end{tabular}
\end{table*}

\subsection{A full kernelized-pair derivative}
\label{subsec:general_functional_validation}

To test the explicit kernel term as well as selection and geometry, use the signed exchange-even kernel
\begin{align}
K_{\bm\beta}(1,2)
=(n_{1z}+n_{2z})e^{0.3u}
\left[1+0.15(x_{c,\delta_1}+x_{d,\delta_2})\right].
\label{eq:benchmark_full_kernel}
\end{align}
Its observed angular arguments are fixed, while its mark derivative is
\begin{align}
\delta K=0.15(n_{1z}+n_{2z})e^{0.3u}(q_c\epsilon_1+q_d\epsilon_2).
\end{align}
We evaluate Eq.~\eqref{eq:exact_boosted_pair_functional} over both angular endpoints and both classes, retaining both solid-angle Jacobians $\delta_1^2\delta_2^2$. Table~\ref{tab:general_validation} reports the derivative of $P/(4\pi)^2$, because the numerical angular weights integrate $\pd\Omega/(4\pi)$ at each endpoint. The five contributions are separately nonzero, so omitting the pair or kernel derivative cannot be hidden by a redefinition of a single coefficient.

\begin{figure*}[t]
\centering
\includegraphics[width=0.73\textwidth]{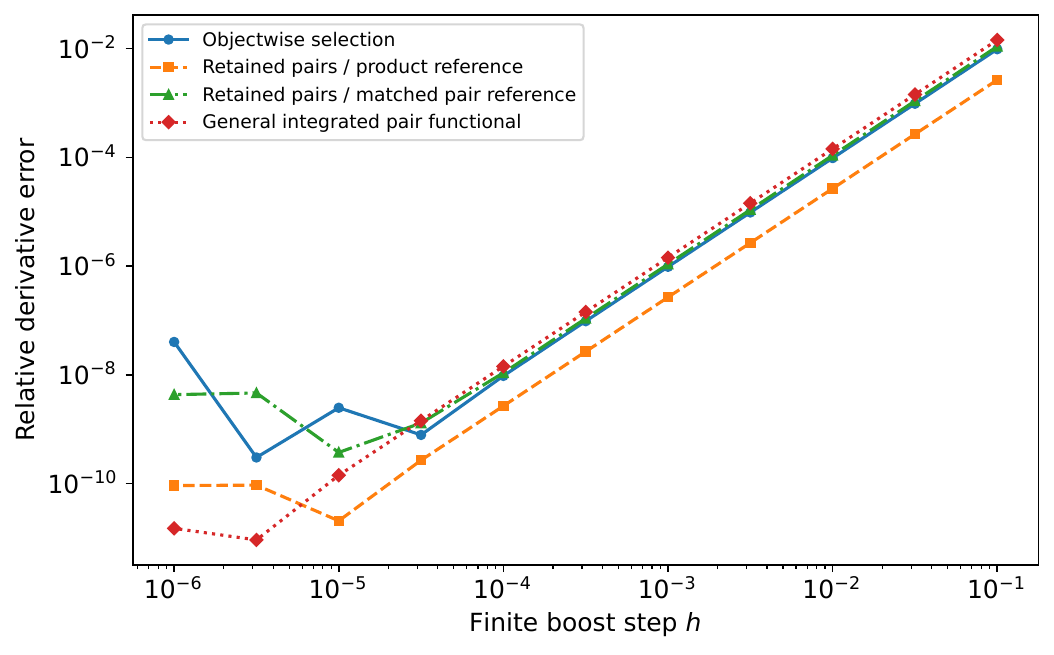}
\caption{Independent finite-boost checks of the objectwise local response, the nonfactorizable retained-pair response under two references, and the full integrated pair functional. Numerical derivatives apply the exact Lorentz map and re-evaluate every selection and normalization factor. The analytic derivatives are assembled from the separate terms. The initial approximately quadratic decrease is the centered-difference truncation error; small-step departures are numerical cancellation, not a survey noise model.}
\label{fig:finite_boost_validation}
\end{figure*}

Quadrature nodes are not catalog objects: coincident angular nodes evaluate a regular factorial density, not a self-pair contact contribution. The default rule uses 12 Gauss--Legendre nodes in $\cos\vartheta$ and 24 equally spaced azimuths per endpoint. A second calculation integrates over fixed parent directions, moves all observed kernel and selection arguments forward, and includes no angular Jacobian in that representation. Its agreement with the observed-coordinate integral checks the transport identity in Eq.~\eqref{eq:angular_transport_identity} without adding the two representations. Increasing these to 16 and 32 changes the analytic total by less than $2\times10^{-14}$ relatively. At $h=10^{-4}$ the finite-boost derivative is $0.3068311286$, compared with $0.3068311242$ from the five-term decomposition. Figure~\ref{fig:finite_boost_validation} shows the centered-difference convergence until floating-point cancellation becomes relevant.
\begin{table}[t]
\centering
\caption{Derivative of the integrated benchmark $P/(4\pi)^2$ along $\hat{\bm v}$. Values follow the five terms in Eq.~\eqref{eq:pair_response_decomposition}.}
\label{tab:general_validation}
\begin{tabular}{lr}
\toprule
Contribution & Derivative\\
\midrule
Angular Jacobians & $0.1618424431$\\
Parent-density aberration & $-0.0010820685$\\
Objectwise selection & $0.0897791538$\\
Pair retention & $0.0366736004$\\
Analysis kernel & $0.0196179955$\\
\midrule
Total & $0.3068311242$\\
\bottomrule
\end{tabular}
\end{table}

\subsection{Additional consistency checks and reproducibility}
\label{subsec:reproducibility_checks}

Independent Legendre quadrature of the configuration-space aberration response reproduces $\ell C_\ell-(\ell+2)C_{\ell+1}$ to a maximum absolute difference below $10^{-14}$ for a finite polynomial test spectrum. Explicit Clebsch--Gordan summation verifies the BipoSH phase and $|\mathcal H_\ell|^2=(\ell+1)/(4\pi)$ to below $10^{-14}$ over the tested low multipoles. Direct spherical-harmonic integration verifies both signs in Eq.~\eqref{eq:contact_neighbor_response}. A finite sample verifies the equality between the self-pair-subtracted harmonic product and the explicit distinct-pair sum. Conditioning a retained-pair count on a Poisson total count independently verifies Eq.~\eqref{eq:retained_poisson_count_check}. These are finite tests of the analytic identities, not proofs for an unbounded numerical implementation.

We also check the two normalization targets in Eq.~\eqref{eq:integrated_midpoint_extra_modulation} and Eq.~\eqref{eq:local_projection_kernel}, the opposite-bias-response populations in Eq.~\eqref{eq:counterexample_opposite_modulations}, the exact proportionality spectrum in Eq.~\eqref{eq:exact_template_degeneracy}, and the fixed-sample hard-cut failure described in Appendix~\ref{appsubsec:forward_model_evaluation}. The last test is especially important: a vanishing finite-sample migration count does not establish a vanishing population response.

The accompanying script \texttt{reproduce\_results.py} defines every parameter, generates the four figures and their numerical tables, and writes machine-readable validation results. Its tests are independent of the manuscript compilation. They establish consistency of the worked expectations and transformations; they do not calibrate masks, non-Gaussian catalog covariance, or a particular observing pipeline. Those remain application-specific tasks.

\section{Connection to Observations and Catalog-Level Tests}
\label{sec:observational_connection}

The preceding results specify quantities that can be measured or calibrated for an observed source catalog. The observational objective is to test whether the count dipole and the allowed two-point anisotropy are consistent with one observer velocity after propagating each sample's selection, reference construction, and discreteness terms. This is a conditional test of a calibrated catalog model, not an assumption that every measured dipole is kinematic. We describe a route to such a test using existing classes of observations; no catalog measurement or survey-specific sensitivity forecast is performed here.

\subsection{Observational targets and the data entering the response}
\label{subsec:observational_catalog_targets}

Radio and mid-infrared catalogs provide concrete starting points. The joint NVSS and RACS analysis of \citet{2024MNRAS.531.4545O} uses angular positions, integrated flux densities, masks, and information identifying nearby radio sources. Its dependence on flux cuts and local-source treatment illustrates why a two-point extension must propagate those choices rather than import a single catalog-wide coefficient. For a radio application, the analysis should also specify whether a catalog entry represents a physical source or one of its resolved components. An association or component-merging rule changes which distinct pairs are available; it belongs to the measurement and selection model, not to a self-pair subtraction. An angular-resolution cut or a calibrated component-association prescription can define a controlled first sample, but the resulting pair selection must still be propagated.

A complementary application is the CatWISE2020 quasar selection used by \citet{2021ApJ...908L..51S}, whose two infrared bands enter catalog construction. Angular clustering has been measured for this population by \citet{2023ApJ...943..116T}. Flux and color variations therefore offer a way to compare the count response, the selected clustering, and the corresponding two-point anisotropy within one observing system. The band responses and color displacement in Eqs.~\eqref{eq:finite_band_response_r4} and \eqref{eq:representative_mark_responses_r4} connect the measured photometry to the selection perturbation. A color does not, by itself, determine a unique spectrum within both passbands; spectral uncertainty must be included in the response calibration. These are proposed applications of the present formalism, not claims that the cited count-dipole or isotropic-clustering analyses have already measured the pair-midpoint signal.

For either example, the required inputs are the selected positions and marks, the actual acceptance and quality rules, the angular coverage and depth information, the pair prescription, and the reference or random-catalog algorithm. Source injection or deeper auxiliary data can constrain the inclusion model. They must describe both the distribution of sources near the selection boundaries and their clustering, because a calibration of marginal completeness alone does not determine the pair-conditioned term. For a pipeline with blending or close-pair losses, joint recovery should be calibrated as a function of angular separation and marks, for example with paired-source injections; two marginal recovery fractions do not establish independent pair inclusion. A denser tomographic imaging sample provides another possible route: \citet{2024JCAP...06..045L} forecast neighboring-multipole velocity inference for a Euclid-like photometric survey. That forecast establishes an observational motivation under its stated assumptions, but its precision cannot be transferred to the radio, infrared, or retained-pair examples considered here without a separate covariance and selection calculation.

\subsection{Measure a specified statistic and calibrate its complete response}
\label{subsec:observational_response_calibration}

The data product should be defined before estimating a velocity. It may be a set of signed pair-midpoint projections in angular bins, or the BipoSH coefficients of Eq.~\eqref{eq:sample_BipoSH_estimator} with the factorial product in Eq.~\eqref{eq:factorial_harmonic_product}. A retained pair list instead requires its explicit distinct-pair sum. In all cases the same mask, reference measure, angular range, mark weights, and self-pair rule must be used in the data and in the prediction. On a partial sky, an equal-area hemispherical contrast or the full-sky factor $1/3$ in Eq.~\eqref{eq:integrated_midpoint_extra_modulation} is not an automatically calibrated velocity estimator.

Let $\widehat d_\alpha$ denote a real component of this specified data vector; $\alpha$ can label a sample, angular bin, or independent harmonic component. Let $\bm\vartheta$ collect the parent-population and instrumental calibration parameters. Define
\begin{align}
\mu_\alpha(\bm\beta,\bm\vartheta)
&\equiv\mathbb E_{\bm\beta,\bm\vartheta}[\widehat d_\alpha],\\
R_{\alpha j}(\bm\vartheta)
&\equiv\left.\frac{\partial\mu_\alpha}{\partial\beta_j}\right|_{\bm\beta=\bm0},
\qquad j=1,2,3,\\
\mu_\alpha(\bm\beta,\bm\vartheta)-\mu_\alpha(\bm0,\bm\vartheta)
&=\sum_{j=1}^{3}R_{\alpha j}(\bm\vartheta)\beta_j+\mathcal O(\beta^2).
\label{eq:observational_response_matrix}
\end{align}
This response includes catalog construction, not only the full-sky harmonic template. The zero-boost mean includes the anisotropy generated by the modeled survey and analysis procedure. In particular, local normalization uses a modeled ensemble intensity, not the realized density of the sky under study. If that intensity is not independently available, one can forward-model the global convention, retaining its extra modulation and contact terms. If mean or nuisance templates are fitted from the data, the fitting procedure must itself be repeated in calibration realizations so that possible signal removal is included in $R$.

An operational estimate of Eq.~\eqref{eq:observational_response_matrix} uses parent realizations with both signs of a small velocity perturbation. Write $X_r(\bm\beta;\bm\vartheta)$ for the selected catalog and reference output of the full observing pipeline for parent realization $r$. With $\hat{\bm e}_j$ a Cartesian unit vector, one can form
\begin{align}
\widehat R_{\alpha j}(h)
={}&\frac{1}{N_{\rm mock}}\sum_{r=1}^{N_{\rm mock}}
\frac{\widehat d_\alpha[X_r(h\hat{\bm e}_j;\bm\vartheta)]}{2h}\nonumber\\*
&-\frac{1}{N_{\rm mock}}\sum_{r=1}^{N_{\rm mock}}
\frac{\widehat d_\alpha[X_r(-h\hat{\bm e}_j;\bm\vartheta)]}{2h}.
\label{eq:observational_response_finite_difference}
\end{align}
The source directions and spectra are transformed before the measurement and final selection are applied. Instrumental coverage and depth remain defined in observed coordinates; their values are evaluated at the transformed source positions. Redshift-dependent cuts and derived photometric quantities are rerun from their transformed inputs. Pair detection or retention, analysis weights, random catalogs, mean fitting, and self-pair terms then follow the declared estimator convention. Paired realizations should share the latent sky and suitably coupled noise or retention random variates. This reduces calibration variance but does not remove it. The range of $h$, the number of parent realizations, and migration across hard boundaries must be checked as in Appendix~\ref{appsubsec:forward_model_evaluation}.

Applying a boost only to an already-selected observed list is not this calibration: it cannot recover objects or pairs that were absent before the perturbation. A deeper parent sample or a validated forward model must cover the relevant boundary neighborhoods. Randomizing marks independently of position is also not a neutral substitute for a clustered parent, because it can erase the score--clustering association being calibrated. Such randomization is useful as a deliberate diagnostic of that association, not as the default physical prediction.

\subsection{Selection variations as an empirical calibration test}
\label{subsec:observational_selection_variations}

The response can be constrained without assigning a noiseless bias value to each individual object. One possibility is to construct nearby flux- or color-selected samples and measure their auto- and cross-clustering with matched survey treatment. A restricted, transparent calibration relation follows from the same rank-one model used in Section~\ref{subsec:linear_bias_selection_model}. For a common scalar flux response $q$, a single threshold, and a fixed underlying projected spectrum,
\begin{align}
C_\ell(S_{\rm lim})&=\bar b(S_{\rm lim})^2C_\ell^{\rm m},\\
\mathcal M_{\rm sel}
&=-\frac{q}{2}\left.
\frac{\partial\ln C_\ell(S_{\rm lim})}{\partial\ln S_{\rm lim}}
\right|_{C_\ell^{\rm m}\,{\rm fixed}}.
\label{eq:observational_threshold_calibration}
\end{align}
Here $C_\ell>0$ is the clustering signal after the appropriate contact and survey corrections. The factor $1/2$ follows from the squared bias, while the sign follows from the threshold displacement $S_{\rm lim}\mapsto S_{\rm lim}e^{-q\varepsilon}$. Equation~\eqref{eq:observational_threshold_calibration} is an operational consequence of Eq.~\eqref{eq:sharp_threshold_bias_correspondence}, not a new contribution in addition to that bias response. It is consistent with the use of selection variations to calibrate response spectra in Appendix~D.3 of \citet{2024JCAP...06..045L}.

A real flux cut can change the redshift distribution and hence the projected matter spectrum, not only the effective bias. In that case differentiating the measured $C_\ell$ is not, by itself, a measurement of $\mathcal M_{\rm sel}$ in Eq.~\eqref{eq:observational_threshold_calibration}. The radial weights must be controlled, or their changes included in the full marked response. Likewise, heterogeneous spectral responses, several simultaneous cuts, nonlinear mark-dependent clustering, and nonfactorizable pair selection require the general score, cross-spectrum, or pair-functional expressions. Where smooth selection scores can be calibrated, mark-binned clustering or a score-weighted map provides access to the signal--response cross-spectrum $S_\ell=C_\ell^{\Delta,\mathcal U}$ in Eq.~\eqref{eq:general_cross_spectral_response}. Its cross-contact contribution and its covariance with the ordinary map must be treated consistently. For hard cuts, the boundary or finite-difference representation is required instead of a pointwise score assigned only to detected objects.

These measurements lead to a testable prediction across selections. Changing the flux or color rule changes the count coefficient, clustering spectrum, response spectrum, and possibly the reference weights. Under the common-rest-frame hypothesis it does not change the observer velocity. Thus it is the \emph{response-corrected velocity}, not the raw count-dipole amplitude or raw midpoint amplitude, that should be consistent across the samples. Joint count-dipole analyses already exploit sample-dependent kinematic expectations \citep{2025A&A...697A.112W}; the present construction specifies the additional information required for the corresponding pair-statistic test. Nested samples and samples on overlapping sky are correlated, even when their object lists are disjoint, and their joint covariance must be propagated rather than treating repeated selections as independent detections.

\subsection{Recovery tests, foreground controls, and cosmological interpretation}
\label{subsec:observational_validation_interpretation}

Before interpreting a data measurement, the pipeline should recover velocities injected into independent validation realizations with both signs and several orientations relative to the survey window. Changing the injection amplitude tests the usable linear range. Separate changes of flux zero point, depth, contamination, nearby-source treatment, and pair resolution test whether nonkinematic perturbations project onto the same data-vector directions. The latter are controls of an assumed measurement model, not effects generated by the observer-boost calculation. Comparing raw-map and factorial statistics, with the contact difference predicted in Section~\ref{subsec:contact_covariance}, and comparing local and integrated normalization conventions provides further internal checks. Purely exchange-odd identical-endpoint pair sums must vanish exactly in the implementation; this algebraic check does not establish the detectability of the allowed midpoint signal.

A practical joint model can be written, to first order around a specified calibration point, as
\begin{align}
\bm d&=\bm R(\bm\vartheta)\bm\beta+\bm T\bm\zeta+\bm\nu,\\
\bm d&\equiv\widehat{\bm d}-\bm\mu(\bm0,\bm\vartheta),
\qquad \mathbb E\bm\nu=\bm0.
\label{eq:observational_joint_model}
\end{align}
The columns of $\bm T$ are explicitly modeled nuisance responses in the same data vector, and $\bm\zeta$ are their small amplitudes. The covariance of $\bm\nu$ includes finite-source sampling, intrinsic large-scale structure under the chosen null ensemble, and measurement noise. Uncertainties in the estimated mean field, response matrix, and calibration parameters must also be propagated, including their correlations with the data when inferred from the same sample. This is an inference prescription, not a universal decomposition of all observed anisotropy into known physical components.

One test fixes $\bm\beta$ to the value inferred from the CMB dipole under its kinematic interpretation and compares the resulting sample-dependent pair predictions with the data. A complementary test fits the velocity components jointly with the specified nuisance parameters. A comparison must use the same observer reference frame in all catalogs. Agreement with the CMB direction alone is insufficient: instrumental patterns can occupy the same $L=1$ sector, and the local-source contribution is material in existing radio count-dipole studies \citep{2024MNRAS.531.4545O}. The constraining evidence is the joint behavior across angular scales, selection variants, normalization conventions, and independent calibration tests. Additional samples help only when their calibrated responses add independent information; the rank limitation in Section~\ref{subsec:identifiability_rank} still applies.

The observational deliverable would therefore be a specified pair measurement together with its measured or simulated response, a validated covariance, and either a calibrated velocity constraint or a limit on identifiable response combinations. A null or low-significance midpoint measurement remains interpretable once this information is supplied. The present four figures establish the consistency and possible consequences of the theory, not the sensitivity of NVSS, RACS, CatWISE2020, or a future survey. Those sensitivities require their actual source densities, measured spectra, selection uncertainties, masks, pair prescriptions, and covariance. The connection to observation is nevertheless explicit: the inputs can be tied to catalog and calibration data, the output is a defined statistic, and the common-velocity prediction can be tested without identifying every observed dipole with observer motion.

\section{Discussion and Conclusions}
\label{sec:conclusion}

We have formulated observer-boost responses for explicitly selected and normalized marked-pair observables. The construction retains the parent pair density, objectwise inclusion, nonfactorizable pair selection, analysis kernel, and reference measure as separate inputs. Its derivative follows from the Lorentz map rather than from a logarithmic derivative of an unspecified count. The equivalence of observed- and rest-coordinate descriptions requires transporting the angular Jacobian and all observed factors consistently.

The selection-dependent modulation of clustering, its distinction from the one-point count dipole, and the aberration--modulation inference problem are established ingredients of previous work, especially \citet{2024JCAP...06..045L}. Our score representation recovers the effective-bias derivative $\partial_\varepsilon\ln\bar b_\varepsilon=\cov(s,b)/\bar b$ and its single-flux-threshold limit. This correspondence identifies the joint population weighting required by a general selected observable and connects it directly to the existing bias-response description. General mark-dependent pair correlations or nonfactorizable pair prescriptions need more than a constant effective-bias coefficient.

For independent objectwise selection, local normalization removes the common endpoint count modulation but leaves a score--pair covariance when clustering and boost sensitivity are associated. Two explicitly admissible populations can have identical one-point response and identical isotropic angular correlation, yet opposite two-point boost response. Nonfactorizable pair retention supplies further terms and makes the choice between a product reference and a retained-pair reference operationally important. The bounded-Cox and pair-list benchmarks test these distinctions using finite boosts rather than repeated substitution into the linear formulas.

A central normalization distinction is that local division and global pair-count normalization do not commute with projection. A signed integrated midpoint statistic can retain $2\mathcal A\cos(\theta/2)w$ even with response-matched randoms. An inverse-intensity kernel or a locally normalized field is required to obtain the local response. Likewise, the self-pair contact term belongs to a raw map product but not to a factorial statistic. We have given explicit sample subtraction and derived the opposite contact-modulation signs for local and global count normalizations. These choices affect the mean template as well as the covariance model.

Endpoint exchange still excludes an ordinary separation-odd same-tracer kernel. The allowed isotropic-reference boost signal is an exchange-even midpoint anisotropy with $L=1$ neighboring-multipole coupling. For general objectwise selection its modulation is described by a signal--response cross-spectrum; a constant $\mathcal M_{\rm sel}$ is a restricted rank-one case. A separately selected pair list remains a pair observable and must not automatically be identified with a single-map covariance.

The toy inference model illustrates both template-mismatch bias and the cost of fitting a selection amplitude. Nearly proportional templates can give a large geometry-only bias while leaving little information for separate velocity inference. Exact proportionality survives any common fixed linear survey operator; adding a nuisance parameter or changing weights cannot create a missing identifiable direction. Spectral structure, genuinely different tracer responses, or external calibration are required. Our Gaussian toy errors are not forecasts for a sparse or masked catalog, for which higher factorial moments and joint retention probabilities must be assessed.

The resulting framework connects existing observer-motion theory to a broader class of catalog-defined pair observables with explicit calibration requirements. Its practical use requires a parent or deeper sample covering relevant selection boundaries, a specified reference construction, an appropriate self-pair convention, and a covariance model matched to the actual pipeline. Preserving these definitions from the source population through the estimator is the main methodological contribution; the independent benchmarks demonstrate how that consistency can be tested.

The observational route in Section~\ref{sec:observational_connection} makes this use concrete: measure a declared pair statistic, calibrate its response through the same selection and normalization pipeline, and compare different catalog selections under a common observer velocity. A velocity constraint or a limit on an identifiable response combination then has a defined observational meaning. Quantifying the signal amplitude and detectability in a specific survey requires dedicated modeling of the source population, selection and pair completeness, survey geometry, and estimator covariance, together with systematic-error control. Survey-specific sensitivity forecasts and applications to observed catalogs are therefore left to future work, building on the theoretical framework and controlled tests presented here.

\begin{appendix}
\renewcommand{\theHequation}{appendix.\Alph{section}.\arabic{equation}}

\section{Selection Boundaries and Practical One-Point Evaluation}
\label{app:selection_response}

The main text defines the one-point kinematic coefficient from the response of the local selected intensity to the direction-dependent Doppler factor. This Appendix develops equivalent boundary, chain-rule, and forward-model representations useful for smooth, sharp, multivariate, and stochastic catalog selections. Throughout, the derivative is taken with respect to $\ln\delta$ at $\delta=1$, not with respect to $\ln\beta$ at $\beta=0$. We retain the parent-variable pullback convention of Sec.~\ref{sec:one_point_boost}: the parent density is held fixed in the reference marks $\bm y$, while the boosted marks $\bm y_\delta$ are inserted into the selection. No additional mark-space Jacobian appears in this representation.

\subsection{Selection response and boost generator}
\label{appsubsec:general_selection_response}

For a statistically isotropic parent population, the unboosted selected intensity at observed direction $\hat{\bm n}$ is
\begin{align}
n_0(\hat{\bm n})
=
\int
f(\bm y)
\mathcal W(\bm y,\hat{\bm n})
\,\pd\bm y.
\label{eq:appendix_unboosted_intensity}
\end{align}
Integration over $\bm y$ includes sums over discrete marks when present. The differential boundary formulas below apply to the continuous mark coordinates; variables that do not transform under the boost have zero generator component.

Let
\begin{align}
G^A(\bm y)
\equiv
\left.
\frac{\partial y_\delta^A}
{\partial\ln\delta}
\right|_{\delta=1}
\label{eq:appendix_mark_generator_components}
\end{align}
be the infinitesimal displacement of the reference marks. If there are $p$ continuous mark coordinates, define
\begin{align}
\mathcal B_{\bm y}
\equiv
\sum_{A=1}^{p}
G^A(\bm y)
\frac{\partial}{\partial y^A}.
\label{eq:appendix_mark_generator}
\end{align}
The angular argument of $\mathcal W$ is held fixed in observed coordinates. Thus, $\mathcal B_{\bm y}$ differentiates source properties and catalog-level derived quantities, whereas the solid-angle Jacobian and angular remapping are treated separately in Sec.~\ref{sec:one_point_boost}.

Define the unnormalized selection response by
\begin{align}
\mathcal S_{\mathcal W}(\hat{\bm n})
&\equiv
\int
f(\bm y)
\mathcal B_{\bm y}
\mathcal W(\bm y,\hat{\bm n})
\,\pd\bm y.
\label{eq:appendix_selection_response_integral}
\end{align}
Where $n_0(\hat{\bm n})>0$, the one-point coefficient is
\begin{align}
\mathcal A[\mathcal W,f](\hat{\bm n})
=
2+
\frac{
\mathcal S_{\mathcal W}(\hat{\bm n})
}{
n_0(\hat{\bm n})
}.
\label{eq:appendix_general_A}
\end{align}
The term $2$ is the solid-angle Jacobian and is not part of the selection derivative. If $\mathcal W$ is smooth in the continuous mark coordinates, then
\begin{align}
\mathcal B_{\bm y}\mathcal W
=
\bm G(\bm y)
\cdot
\nabla_{\bm y}\mathcal W.
\label{eq:appendix_smooth_generator_action}
\end{align}
For a discontinuous selection, Eq.~\eqref{eq:appendix_smooth_generator_action} is understood weakly. The integral in Eq.~\eqref{eq:appendix_selection_response_integral} is invariant under a smooth reparameterization of the continuous marks, because $\mathcal W$ is a scalar, $\mathcal B_{\bm y}$ is a vector field, and $f(\bm y)\,\pd\bm y$ is the parent measure.

\subsection{Deterministic multivariate boundaries}
\label{appsubsec:deterministic_boundaries}

Suppose that, at fixed $\hat{\bm n}$, the selected region is defined by $J$ inequalities,
\begin{align}
\mathcal D(\hat{\bm n})
=
\left\{
\bm y:
F_j(\bm y,\hat{\bm n})\geq0,
\quad
j=1,\ldots,J
\right\},
\end{align}
so that
\begin{align}
\mathcal W(\bm y,\hat{\bm n})
=
\prod_{j=1}^{J}
\Theta
\left[
F_j(\bm y,\hat{\bm n})
\right].
\label{eq:appendix_multiboundary_selection}
\end{align}
The distributional boost derivative is
\begin{align}
\mathcal B_{\bm y}\mathcal W
={}&
\sum_{j=1}^{J}
\deltadir
\left[
F_j(\bm y,\hat{\bm n})
\right]
\mathcal B_{\bm y}F_j(\bm y,\hat{\bm n})
\nonumber\\*
&\times
\prod_{k\neq j}
\Theta
\left[
F_k(\bm y,\hat{\bm n})
\right].
\label{eq:appendix_multiboundary_derivative}
\end{align}
Assume distinct active boundary facets and that each is regular,
\begin{align}
\left|
\nabla_{\bm y}F_j
\right|
\neq0
\qquad
\text{on }F_j=0.
\end{align}
The coarea formula then gives
\begin{align}
\mathcal S_{\mathcal W}(\hat{\bm n})
={}&
\sum_{j=1}^{J}
\int_{F_j=0}
\frac{
f(\bm y)
\mathcal B_{\bm y}F_j(\bm y,\hat{\bm n})
}{
\left|
\nabla_{\bm y}F_j(\bm y,\hat{\bm n})
\right|
}
\nonumber\\*
&\times
\prod_{k\neq j}
\Theta
\left[
F_k(\bm y,\hat{\bm n})
\right]
\,\pd\Sigma_j.
\label{eq:appendix_boundary_surface_integral}
\end{align}
The factor
\begin{align}
v_{\perp,j}(\bm y,\hat{\bm n})
\equiv
\frac{
\mathcal B_{\bm y}F_j(\bm y,\hat{\bm n})
}{
\left|
\nabla_{\bm y}F_j(\bm y,\hat{\bm n})
\right|
}
\end{align}
is the component of the mark displacement toward increasing $F_j$, normal to the boundary. A positive value carries parent objects into the selected region $F_j\geq0$; the location of the pulled-back boundary in fixed parent coordinates moves in the opposite direction. Equation~\eqref{eq:appendix_boundary_surface_integral} is therefore the signed population flux, without a sign ambiguity between object motion and boundary motion. We assume distinct regular facets whose intersections have zero induced surface measure, for example transverse intersections, and no singular parent mass on them. Under these assumptions the intersections add no first-order term. Coincident or redundant cuts must first be reduced to a single indicator or treated through a specified smooth limit; products such as a Dirac distribution with a discontinuous indicator at the same boundary are not independently defined by the formal product rule. Singular parent components require a separate measure-level treatment.

The one-dimensional sharp-threshold limit follows immediately by taking $F=S-S_{\rm lim}$ in Eq.~\eqref{eq:appendix_boundary_surface_integral}; it reproduces Eq.~\eqref{eq:threshold_generalized_EB} without requiring a catalog-wide power law.

\subsection{Smooth probabilistic selection}
\label{appsubsec:smooth_selection}

For a smooth one-dimensional completeness function $\mathcal W(S,\hat{\bm n})$, with $\hat{\bm n}$ held fixed, the intensity and selection response are
\begin{align}
n_0(\hat{\bm n})
&=
\int
f(S,\bm\eta)
\mathcal W(S,\hat{\bm n})
\,\pd S\pd\bm\eta,
\nonumber\\*
\mathcal S_{\mathcal W}(\hat{\bm n})
&=
\int
f(S,\bm\eta)
q(S,\bm\eta)S
\frac{
\partial\mathcal W(S,\hat{\bm n})
}{
\partial S
}
\,\pd S\pd\bm\eta.
\label{eq:appendix_smooth_flux_response}
\end{align}
Therefore,
\begin{align}
\mathcal A(\hat{\bm n})
={}&
2+
\frac{1}{n_0(\hat{\bm n})}
\int
f(S,\bm\eta)
q(S,\bm\eta)S
\nonumber\\*
&\times
\frac{
\partial\mathcal W(S,\hat{\bm n})
}{
\partial S
}
\,\pd S\pd\bm\eta.
\label{eq:appendix_smooth_A}
\end{align}
The response is distributed across the transition region in which the inclusion probability changes. If
\begin{align}
\left[
\mathcal W(S,\hat{\bm n})
f(S,\bm\eta)
q(S,\bm\eta)S
\right]_{S=0}^{S=\infty}
=
0
\end{align}
after integration over $\bm\eta$, integration by parts gives
\begin{align}
\mathcal S_{\mathcal W}(\hat{\bm n})
=
-
\int
\mathcal W(S,\hat{\bm n})
\frac{\partial}{\partial S}
\left[
f(S,\bm\eta)
q(S,\bm\eta)S
\right]
\,\pd S\pd\bm\eta.
\label{eq:appendix_smooth_integration_by_parts}
\end{align}
This form avoids differentiating a numerically tabulated completeness curve, but requires a differentiable model for the population-weighted flux response.

\subsection{General multivariate and stochastic selection}
\label{appsubsec:multivariate_selection}

Suppose that catalog membership depends on measured or derived variables
\begin{align}
\bm u(\bm y,\hat{\bm n})
=
\left(
u_1(\bm y,\hat{\bm n}),\ldots,
u_m(\bm y,\hat{\bm n})
\right)
\end{align}
through
\begin{align}
\mathcal W(\bm y,\hat{\bm n})
=
\widetilde{\mathcal W}
\left[
\bm u(\bm y,\hat{\bm n}),
\hat{\bm n}
\right].
\end{align}
Define
\begin{align}
Q_k(\bm y,\hat{\bm n})
&\equiv
\left.
\frac{\partial u_{k,\delta}}
{\partial\ln\delta}
\right|_{\delta=1}
\nonumber\\*
&=
\mathcal B_{\bm y}
u_k(\bm y,\hat{\bm n}).
\label{eq:appendix_derived_variable_response}
\end{align}
The chain rule gives
\begin{align}
\mathcal B_{\bm y}\mathcal W
=
\sum_{k=1}^{m}
\frac{
\partial\widetilde{\mathcal W}
}{
\partial u_k
}
Q_k(\bm y,\hat{\bm n}),
\label{eq:appendix_multivariate_chain_rule}
\end{align}
and hence
\begin{align}
\mathcal A[\mathcal W,f](\hat{\bm n})
={}&
2+
\frac{1}{n_0(\hat{\bm n})}
\int
f(\bm y)
\nonumber\\*
&\times
\sum_{k=1}^{m}
\frac{
\partial\widetilde{\mathcal W}
}{
\partial u_k
}
\nonumber\\*
&\qquad\times
Q_k(\bm y,\hat{\bm n})
\,\pd\bm y.
\label{eq:appendix_multivariate_A}
\end{align}
The variables $u_k$ may include band fluxes, colors, photometric-redshift summaries, morphology indicators, or classification scores. Their responses must be propagated through the same measurement and catalog pipeline used to define the sample. A fixed angular mask or depth map remains in the explicit $\hat{\bm n}$ dependence of $\widetilde{\mathcal W}$; its angular mode coupling is not part of the mark derivative $\mathcal B_{\bm y}$. For hard multivariate cuts, Eq.~\eqref{eq:appendix_multivariate_chain_rule} is replaced by the boundary form in Eq.~\eqref{eq:appendix_multiboundary_derivative}.

A noisy or stochastic measurement pipeline can be included by introducing the conditional distribution of measured variables $\bm u$ at fixed parent marks,
\begin{align}
\mathcal W_\delta(\bm y,\hat{\bm n})
=
\int
\Pi_{\rm cat}(\bm u,\hat{\bm n})
p_\delta(\bm u\mid\bm y,\hat{\bm n})
\,\pd\bm u,
\label{eq:appendix_stochastic_selection}
\end{align}
where $0\leq\Pi_{\rm cat}\leq1$ is the final acceptance rule or inclusion probability. Its response is
\begin{align}
\dot{\mathcal W}(\bm y,\hat{\bm n})
=
\int
\Pi_{\rm cat}(\bm u,\hat{\bm n})
\left.
\frac{\partial p_\delta(\bm u\mid\bm y,\hat{\bm n})}
{\partial\ln\delta}
\right|_{\delta=1}
\,\pd\bm u.
\label{eq:appendix_stochastic_selection_response}
\end{align}
A deterministic measurement map is recovered when $p_\delta$ is a Dirac measure concentrated on $\bm u(\bm y_\delta,\hat{\bm n})$. Equations~\eqref{eq:appendix_stochastic_selection} and \eqref{eq:appendix_stochastic_selection_response} show that flux migration, measurement scatter, and probabilistic classification should be perturbed before the final catalog cut is applied.

\subsection{Forward-model and finite-sample evaluation}
\label{appsubsec:forward_model_evaluation}

The selection derivative can be evaluated without differentiating the catalog boundary analytically. Introduce a scalar perturbation $\varepsilon=\ln\delta$ and define the selection-only intensity
\begin{align}
\mathcal I_{\mathcal W}(\varepsilon,\hat{\bm n})
\equiv
\int
f(\bm y)
\mathcal W
\left(
\bm y_{\exp(\varepsilon)},
\hat{\bm n}
\right)
\,\pd\bm y.
\label{eq:appendix_boosted_selection_integral}
\end{align}
Then
\begin{align}
\mathcal I_{\mathcal W}(0,\hat{\bm n})
&=
n_0(\hat{\bm n}),
\nonumber\\*
\left.
\frac{\partial\mathcal I_{\mathcal W}}
{\partial\varepsilon}
\right|_{\varepsilon=0}
&=
\mathcal S_{\mathcal W}(\hat{\bm n}),
\end{align}
and
\begin{align}
\mathcal A(\hat{\bm n})
=
2+
\left.
\frac{\partial\ln\mathcal I_{\mathcal W}(\varepsilon,\hat{\bm n})}
{\partial\varepsilon}
\right|_{\varepsilon=0}.
\end{align}
For sufficiently small $h$,
\begin{align}
\mathcal A(\hat{\bm n})
={}&
2+
\frac{
\ln\mathcal I_{\mathcal W}(h,\hat{\bm n})
-
\ln\mathcal I_{\mathcal W}(-h,\hat{\bm n})
}{
2h
}
+
\mathcal O(h^2).
\label{eq:appendix_central_difference_response}
\end{align}
This form applies equally to smooth and hard selections, provided that the population integral is sufficiently regular.

For a weighted parent or proposal sample $\{\bm y_i,w_i\}_{i=1}^{N}$ that covers both sides of every relevant selection boundary, define
\begin{align}
\widehat{\mathcal I}_{\mathcal W}(\varepsilon,\hat{\bm n})
\equiv
\sum_{i=1}^{N}
w_i
\mathcal W
\left(
\bm y_{i,\exp(\varepsilon)},
\hat{\bm n}
\right).
\end{align}
For a smooth selection, the direct estimator is
\begin{align}
\widehat{\mathcal A}(\hat{\bm n})
=
2+
\frac{
\displaystyle
\sum_{i=1}^{N}
w_i
\mathcal B_{\bm y_i}
\mathcal W(\bm y_i,\hat{\bm n})
}{
\displaystyle
\sum_{i=1}^{N}
w_i
\mathcal W(\bm y_i,\hat{\bm n})
}.
\label{eq:appendix_weighted_sample_response}
\end{align}
For a hard or pipeline-defined selection, define the finite-sample statistic, wherever both intensity estimates are positive, by
\begin{align}
\widehat{\mathcal A}_{N,h}(\hat{\bm n})
\equiv2+\frac{\ln\widehat{\mathcal I}_{\mathcal W}(h,\hat{\bm n})-\ln\widehat{\mathcal I}_{\mathcal W}(-h,\hat{\bm n})}{2h}.
\label{eq:appendix_weighted_finite_difference}
\end{align}
No deterministic $\mathcal O(h^2)$ remainder is attached to this definition. The population centered difference $\mathcal A_h$ has $\mathcal A_h-\mathcal A=\mathcal O(h^2)$ when the logarithm of the population integral is three times differentiable. The empirical error $\widehat{\mathcal A}_{N,h}-\mathcal A_h$ is a separate sampling term. For hard cuts and fixed $N$, $\widehat{\mathcal I}(\varepsilon)$ is generally a step function. Taking $h\to0$ then resolves no crossings and can make the estimated selection response zero even when the population response is nonzero.

For example, take a uniform parent on $[-1,1]$ and selection $\Theta(y+\varepsilon)$. Its population intensity is $(1+\varepsilon)/2$ near zero, so $\mathcal A=3$. For the fixed sample $(-0.8,-0.2,0.15,0.5)$ and $0<h<0.15$, both perturbed catalogs contain two objects and Eq.~\eqref{eq:appendix_weighted_finite_difference} gives $\widehat{\mathcal A}_{N,h}=2$. The discrepancy is unresolved boundary migration, not a failure of the population derivative.

Under an independent proposal sample with bounded weights, positive baseline intensity, regular nonzero boundary-crossing density, and common random numbers, the numerator difference of a hard-cut intensity estimate has variance of order $h/N$. Its derivative consequently has a leading sampling scale of order $(Nh)^{-1/2}$. Under these assumptions, a representative joint limit is
\begin{align}
h&\to0,\qquad Nh\to\infty,\nonumber\\*
\widehat{\mathcal A}_{N,h}-\mathcal A
&=\mathcal O(h^2)+\mathcal O_p[(Nh)^{-1/2}].
\label{eq:hard_cut_joint_limit_r4}
\end{align}
This rate is not universal: dependent samples, unbounded importance weights, singular boundaries, or a different coupling of stochastic measurements can change it. The ratio and logarithm also produce finite-sample bias; positivity and moment conditions are needed for a delta-method approximation. For smooth inclusion probabilities, direct differentiation as in Eq.~\eqref{eq:appendix_weighted_sample_response} avoids the unresolved-crossing limit and ordinarily has the usual Monte Carlo scale under finite-variance assumptions.

The normalization of the weights cancels. Use the same parent objects and, for stochastic pipelines, coupled random variates for the $+h$ and $-h$ evaluations. This reduces variance but does not remove it. Check a range of step sizes and sample sizes, report crossing counts or effective boundary sample sizes for hard cuts, and separate derivative truncation from sampling and numerical errors. Differentiating an expected inclusion probability is different from differentiating one finite realization of Bernoulli acceptance decisions.

A catalog containing only already-selected objects is generally insufficient for Eqs.~\eqref{eq:appendix_weighted_sample_response} and \eqref{eq:appendix_weighted_finite_difference}, because objects initially outside the catalog can enter after the perturbation. Practical evaluation therefore requires a parent sample, a deeper auxiliary catalog, or a forward simulation whose support includes the relevant boundary neighborhoods. Direction-dependent selection can be handled by evaluating the same construction in angular pixels or by passing each simulated object through the appropriate local survey model.

\subsection{Interpretation and relation to pair statistics}
\label{appsubsec:selection_response_interpretation}

Equations~\eqref{eq:appendix_boundary_surface_integral}, \eqref{eq:appendix_smooth_A}, \eqref{eq:appendix_multivariate_A}, and \eqref{eq:appendix_central_difference_response} give complementary representations of the same one-point endpoint response. The catalog-dependent part of $\mathcal A$ is controlled by the displacement of hard selection boundaries, by the gradient of a smooth inclusion probability, or by the finite-difference response of the complete forward model. Realistic bandpasses, heterogeneous spectra, colors, redshift-dependent selections, measurement scatter, and probabilistic completeness change both the boost displacement and the population weighting of these boundaries \citep{2026ApJS..285....3T}.

For tracer $a$, the normalized one-point selection score used in Sec.~\ref{sec:two_point_boost_response} is
\begin{align}
\sigma_a
=
\mathcal A_a-2
=
\frac{
\mathcal S_{\mathcal W_a}
}{
n_a
}.
\end{align}
It enters the disconnected pair normalization, but it is not generally equal to the corresponding pair-conditioned score in Eq.~\eqref{eq:first_pair_conditioned_selection_score}. Their difference, together with pair-completeness response, forms $\Delta\Sigma_{ab}^{[1]}$ and $\Delta\Sigma_{ab}^{[2]}$ in Eq.~\eqref{eq:pair_selection_residuals}. For a sharp endpoint selection, the pair-conditioned numerator is obtained from the same boundary distribution as Eq.~\eqref{eq:appendix_multiboundary_derivative}, weighted by the selected second factorial-moment density rather than by the one-point parent density. This distinction is why a one-point Ellis--Baldwin coefficient alone does not determine the response of a normalized two-point statistic.

\section{Pair-Geometry and Normalization Details}
\label{app:pair_response}
\label{app:correlation_dipole_derivation}

This Appendix derives the local angular two-point response directly from the Lorentz transformation of the two endpoints. We use the directional variation $\delta_{\bm\beta}$ defined in Eq.~\eqref{eq:directional_boost_variation} and the observed-coordinate pullback convention of Sec.~\ref{sec:two_point_boost_response}. No independent transformation law is assigned to a scalar pair separation. Instead, aberration is applied to each observed direction, the selected second factorial-moment density is transformed, and the response of the one-point normalization is subtracted. The result is the exchange-even pair-midpoint anisotropy used in the main text. A separation-odd channel appears only when the observable contains tracer, mark, or endpoint ordering. Irreducible pair-completeness responses, boost-dependent analysis weights, and survey-window terms remain covered by the explicit pair-functional derivatives in Sec.~\ref{subsec:first_order_pair_decomposition}; the local derivation below isolates the terms that close on an isotropic angular correlation.

\subsection{Aberration of an angular pair}
\label{appsubsec:angular_pair_aberration}

Let $\hat{\bm n}_1$ and $\hat{\bm n}_2$ be the two observed directions. Away from the coincident and antipodal configurations, define
\begin{align}
u
&\equiv
\hat{\bm n}_1\cdot\hat{\bm n}_2
=
\cos\theta,
\nonumber\\*
\delta_i
&\equiv
\delta(\hat{\bm n}_i),
\qquad
\epsilon_i
\equiv
\bm\beta\cdot\hat{\bm n}_i,
\nonumber\\*
\bm\beta_{\perp i}
&\equiv
\bm\beta-
\epsilon_i\hat{\bm n}_i,
\qquad
i=1,2.
\label{eq:appb_pair_variables}
\end{align}
At fixed observed directions, the corresponding cosmic-rest-frame directions are
\begin{align}
\hat{\bm n}_i'
=
\hat{\bm n}_i-
\bm\beta_{\perp i}
+
\mathcal O(\beta^2).
\label{eq:appb_endpoint_aberration}
\end{align}
Their scalar product is therefore
\begin{align}
u'
&=
\hat{\bm n}_1'\cdot\hat{\bm n}_2'
\nonumber\\*
&=
u-
\bm\beta_{\perp1}\cdot\hat{\bm n}_2-
\bm\beta_{\perp2}\cdot\hat{\bm n}_1
+
\mathcal O(\beta^2)
\nonumber\\*
&=
u-
(1-u)
(\epsilon_1+\epsilon_2)
+
\mathcal O(\beta^2).
\label{eq:appb_aberrated_pair_cosine}
\end{align}
Since $\pd u=-\sin\theta\,\pd\theta$, Eq.~\eqref{eq:appb_aberrated_pair_cosine} gives
\begin{align}
\theta'
=
\theta+
\tan\left(
\frac{\theta}{2}
\right)
(\epsilon_1+\epsilon_2)
+
\mathcal O(\beta^2).
\label{eq:appb_aberrated_pair_angle}
\end{align}

The exchange-even midpoint direction and exchange-odd oriented-separation direction are
\begin{align}
\hat{\bm N}
&\equiv
\frac{
\hat{\bm n}_1+\hat{\bm n}_2
}{
2\cos(\theta/2)
},
\nonumber\\*
\hat{\bm s}
&\equiv
\frac{
\hat{\bm n}_1-\hat{\bm n}_2
}{
2\sin(\theta/2)
}.
\label{eq:appb_midpoint_separation_directions}
\end{align}
They satisfy
\begin{align}
\epsilon_1+\epsilon_2
&=
2\cos\left(
\frac{\theta}{2}
\right)
\bm\beta\cdot\hat{\bm N},
\nonumber\\*
\epsilon_1-\epsilon_2
&=
2\sin\left(
\frac{\theta}{2}
\right)
\bm\beta\cdot\hat{\bm s}.
\label{eq:appb_endpoint_sum_difference}
\end{align}
Substitution into Eq.~\eqref{eq:appb_aberrated_pair_angle} yields
\begin{align}
\theta'
=
\theta+
2\sin\left(
\frac{\theta}{2}
\right)
\bm\beta\cdot\hat{\bm N}
+
\mathcal O(\beta^2).
\label{eq:appb_midpoint_angle_shift}
\end{align}
Thus, aberration of the angular separation is controlled by the orientation of the pair midpoint. It contains no term proportional to $\bm\beta\cdot\hat{\bm s}$. At the antipodal configuration $\hat{\bm N}$ is undefined, but the regular combination $(\hat{\bm n}_1+\hat{\bm n}_2)F(u)$ vanishes there whenever the response is nonsingular.

\subsection{Pure geometrical response of an isotropic correlation}
\label{appsubsec:pure_geometrical_response}

Consider a locally normalized angular correlation $w_0(u)$ in the pair-complete, mark-separable limit. In the observed-coordinate representation, the moving observer evaluates the rest-frame correlation at the remapped geometry:
\begin{align}
 w_{\bm\beta}
\left(
\hat{\bm n}_1,
\hat{\bm n}_2
\right)
=
 w_0(u')
+
\mathcal O(\beta^2).
\label{eq:appb_pure_aberration_pullback}
\end{align}
Using Eq.~\eqref{eq:appb_aberrated_pair_cosine}, its first-order variation is
\begin{align}
\delta_{\bm\beta}w
&=
-(1-u)
(\epsilon_1+\epsilon_2)
\frac{\partial w_0}{\partial u},
\label{eq:appb_pure_response_u}
\\
&=
\tan\left(
\frac{\theta}{2}
\right)
(\epsilon_1+\epsilon_2)
\frac{\partial w_0}{\partial\theta},
\label{eq:appb_pure_response_theta_intermediate}
\\
&=
2\sin\left(
\frac{\theta}{2}
\right)
\bm\beta\cdot\hat{\bm N}
\frac{\partial w_0}{\partial\theta}.
\label{eq:appb_pure_midpoint_response}
\end{align}
Equation~\eqref{eq:appb_pure_midpoint_response} reproduces Eq.~\eqref{eq:auto_midpoint_geometrical_limit}.

The same angular structure follows without using the explicit aberration map. For a parity-even isotropic reference ensemble, the most general scalar correction linear in $\bm\beta$ is
\begin{align}
\delta_{\bm\beta}w
={}&
\bm\beta\cdot
\left(
\hat{\bm n}_1+\hat{\bm n}_2
\right)
F^{(+)}(u)
\nonumber\\*
&+
\bm\beta\cdot
\left(
\hat{\bm n}_1-\hat{\bm n}_2
\right)
F^{(-)}(u).
\label{eq:appb_general_linear_pair_scalar}
\end{align}
For a same-tracer auto-correlation with identical endpoint treatment,
\begin{align}
 w_{\bm\beta}
\left(
\hat{\bm n}_1,
\hat{\bm n}_2
\right)
=
 w_{\bm\beta}
\left(
\hat{\bm n}_2,
\hat{\bm n}_1
\right),
\end{align}
which requires
\begin{align}
F^{(-)}(u)=0.
\label{eq:appb_auto_separation_term_zero}
\end{align}
The logarithmic-slope term in Eq.~\eqref{eq:appb_pure_midpoint_response} is therefore an exchange-even midpoint modulation, not an ordinary same-tracer separation dipole.

\subsection{Normalized response with selection and pair completeness}
\label{appsubsec:normalized_selected_response}

We next derive the local selection terms used in Sec.~\ref{sec:two_point_boost_response}. We assume a statistically isotropic rest-frame parent population, no explicit angular dependence in the objectwise selections, and a pair-completeness factor whose geometrical arguments are defined in observed coordinates. We further assume that the first-order pair-completeness response is generated by the transformed endpoint marks and can be decomposed into endpoint contributions. A boost-dependent analysis kernel, an irreducible pair-level response, or a survey window is instead retained in the general response terms of Eq.~\eqref{eq:pair_response_decomposition}.

Under these assumptions, the selected angular pair density is
\begin{align}
\mathcal N_{ab,\bm\beta}^{(2)}
\left(
\hat{\bm n}_1,
\hat{\bm n}_2
\right)
={}&
\delta_1^2\delta_2^2
\int
\rho_{2,ab}'
\left(
 u',
\bm y_1,
\bm y_2
\right)
\nonumber\\*
&\times
\mathcal W_a
\left(
\bm y_{1,\delta_1}
\right)
\mathcal W_b
\left(
\bm y_{2,\delta_2}
\right)
\nonumber\\*
&\times
\mathcal C_{ab,\bm\beta}
\left(
 u,
\bm y_1,
\bm y_2
\right)
\,\pd\bm y_1\pd\bm y_2.
\label{eq:appb_boosted_local_pair_density}
\end{align}
The corresponding one-point intensities are
\begin{align}
 n_{a,\bm\beta}(\hat{\bm n}_1)
&=
\delta_1^2
\int
\rho_{1,a}'(\bm y_1)
\mathcal W_a
\left(
\bm y_{1,\delta_1}
\right)
\,\pd\bm y_1,
\nonumber\\*
 n_{b,\bm\beta}(\hat{\bm n}_2)
&=
\delta_2^2
\int
\rho_{1,b}'(\bm y_2)
\mathcal W_b
\left(
\bm y_{2,\delta_2}
\right)
\,\pd\bm y_2.
\label{eq:appb_boosted_local_one_point_intensities}
\end{align}
The locally normalized correlation is defined by
\begin{align}
1+w_{ab,\bm\beta}
\left(
\hat{\bm n}_1,
\hat{\bm n}_2
\right)
=
\frac{
\mathcal N_{ab,\bm\beta}^{(2)}
\left(
\hat{\bm n}_1,
\hat{\bm n}_2
\right)
}{
 n_{a,\bm\beta}(\hat{\bm n}_1)
 n_{b,\bm\beta}(\hat{\bm n}_2)
}.
\label{eq:appb_boosted_normalized_correlation}
\end{align}

At $\bm\beta=\bm 0$, write
\begin{align}
\mathcal N_{ab}^{(2)}(u)
={}&
\int
\rho_{2,ab}'
\left(
 u,
\bm y_1,
\bm y_2
\right)
\mathcal W_a(\bm y_1)
\mathcal W_b(\bm y_2)
\nonumber\\*
&\times
\mathcal C_{ab}
\left(
 u,
\bm y_1,
\bm y_2
\right)
\,\pd\bm y_1\pd\bm y_2,
\label{eq:appb_unboosted_local_pair_density}
\\
1+w_{ab}(u)
&=
\frac{
\mathcal N_{ab}^{(2)}(u)
}{
 n_an_b
}.
\label{eq:appb_unboosted_normalized_correlation}
\end{align}
Define the one-point selection scores
\begin{align}
\sigma_a
&\equiv
\frac{1}{n_a}
\int
\rho_{1,a}'(\bm y_1)
\dot{\mathcal W}_a(\bm y_1)
\,\pd\bm y_1,
\nonumber\\*
\sigma_b
&\equiv
\frac{1}{n_b}
\int
\rho_{1,b}'(\bm y_2)
\dot{\mathcal W}_b(\bm y_2)
\,\pd\bm y_2,
\label{eq:appb_one_point_selection_scores}
\end{align}
where
\begin{align}
\dot{\mathcal W}_a
&\equiv
\left.
\frac{\partial}{\partial\ln\delta_1}
\mathcal W_a
\left(
\bm y_{1,\delta_1}
\right)
\right|_{\delta_1=1}
=
\mathcal B_{\bm y_1}\mathcal W_a,
\nonumber\\*
\dot{\mathcal W}_b
&\equiv
\left.
\frac{\partial}{\partial\ln\delta_2}
\mathcal W_b
\left(
\bm y_{2,\delta_2}
\right)
\right|_{\delta_2=1}
=
\mathcal B_{\bm y_2}\mathcal W_b.
\end{align}
For a hard selection, these derivatives are understood weakly and are supported on the selection boundaries described in Appendix~\ref{app:selection_response}.

The pair-conditioned endpoint scores are
\begin{align}
\sigma_{a\mid ab}^{[1]}(u)
\equiv{}&
\frac{1}{\mathcal N_{ab}^{(2)}(u)}
\int
\rho_{2,ab}'
\dot{\mathcal W}_a
\mathcal W_b
\mathcal C_{ab}
\,\pd\bm y_1\pd\bm y_2,
\nonumber\\*
\sigma_{b\mid ab}^{[2]}(u)
\equiv{}&
\frac{1}{\mathcal N_{ab}^{(2)}(u)}
\int
\rho_{2,ab}'
\mathcal W_a
\dot{\mathcal W}_b
\mathcal C_{ab}
\,\pd\bm y_1\pd\bm y_2.
\label{eq:appb_pair_conditioned_selection_scores}
\end{align}
Common arguments $(u,\bm y_1,\bm y_2)$ have been suppressed in these and the following integrands. If the pair-completeness response is generated by the two transformed endpoint marks, write
\begin{align}
\delta_{\bm\beta}\mathcal C_{ab}
=
\epsilon_1
\dot{\mathcal C}_{ab}^{[1]}
+
\epsilon_2
\dot{\mathcal C}_{ab}^{[2]},
\qquad
\dot{\mathcal C}_{ab}^{[i]}
\equiv
\mathcal B_{\bm y_i}\mathcal C_{ab}.
\end{align}
The normalized pair-completeness scores are then
\begin{align}
\kappa_{ab}^{[1]}(u)
\equiv{}&
\frac{1}{\mathcal N_{ab}^{(2)}(u)}
\int
\rho_{2,ab}'
\mathcal W_a
\mathcal W_b
\dot{\mathcal C}_{ab}^{[1]}
\,\pd\bm y_1\pd\bm y_2,
\nonumber\\*
\kappa_{ab}^{[2]}(u)
\equiv{}&
\frac{1}{\mathcal N_{ab}^{(2)}(u)}
\int
\rho_{2,ab}'
\mathcal W_a
\mathcal W_b
\dot{\mathcal C}_{ab}^{[2]}
\,\pd\bm y_1\pd\bm y_2.
\label{eq:appb_pair_completeness_scores}
\end{align}
Finally, define the geometrical score
\begin{align}
\mathcal G_{ab}(u)
\equiv
-
\frac{
1-u
}{
\mathcal N_{ab}^{(2)}(u)
}
\int
\frac{\partial\rho_{2,ab}'}{\partial u}
\mathcal W_a
\mathcal W_b
\mathcal C_{ab}
\,\pd\bm y_1\pd\bm y_2.
\label{eq:appb_geometrical_score}
\end{align}

Expanding Eq.~\eqref{eq:appb_boosted_local_pair_density} gives
\begin{align}
\frac{
\delta_{\bm\beta}\mathcal N_{ab}^{(2)}
}{
\mathcal N_{ab}^{(2)}
}
={}&
2(\epsilon_1+\epsilon_2)
+
(\epsilon_1+\epsilon_2)
\mathcal G_{ab}
\nonumber\\*
&+
\epsilon_1
\left(
\sigma_{a\mid ab}^{[1]}
+
\kappa_{ab}^{[1]}
\right)
\nonumber\\*
&+
\epsilon_2
\left(
\sigma_{b\mid ab}^{[2]}
+
\kappa_{ab}^{[2]}
\right).
\label{eq:appb_pair_density_log_response}
\end{align}
The one-point responses are
\begin{align}
\frac{
\delta_{\bm\beta}n_a
}{
 n_a
}
&=
(2+\sigma_a)\epsilon_1,
\nonumber\\*
\frac{
\delta_{\bm\beta}n_b
}{
 n_b
}
&=
(2+\sigma_b)\epsilon_2.
\label{eq:appb_one_point_log_responses}
\end{align}
The two universal solid-angle Jacobians cancel when Eq.~\eqref{eq:appb_boosted_normalized_correlation} is differentiated. Defining
\begin{align}
\Delta\Sigma_{ab}^{[1]}(u)
&\equiv
\sigma_{a\mid ab}^{[1]}(u)
+
\kappa_{ab}^{[1]}(u)
-
\sigma_a,
\nonumber\\*
\Delta\Sigma_{ab}^{[2]}(u)
&\equiv
\sigma_{b\mid ab}^{[2]}(u)
+
\kappa_{ab}^{[2]}(u)
-
\sigma_b,
\label{eq:appb_residual_endpoint_scores}
\end{align}
one obtains
\begin{align}
\delta_{\bm\beta}w_{ab}
={}&
\left[
1+w_{ab}(u)
\right]
\Bigl\{
(\epsilon_1+\epsilon_2)
\mathcal G_{ab}(u)
\nonumber\\*
&+
\epsilon_1
\Delta\Sigma_{ab}^{[1]}(u)
+
\epsilon_2
\Delta\Sigma_{ab}^{[2]}(u)
\Bigr\}.
\label{eq:appb_normalized_response_master}
\end{align}
This is the detailed derivation of Eq.~\eqref{eq:local_angular_response_master}. It describes division by the two boosted means at each pair, not a ratio of already-integrated pair counts. Section~\ref{sec:normalization_contact} constructs the inverse-intensity kernel that measures this local quantity. An integrated signed projection can retain one-point density modulation even with a response-matched random catalog, as Eq.~\eqref{eq:integrated_midpoint_extra_modulation} demonstrates. The associated scalar quotient rule does not by itself reproduce this local response.

\subsection{Midpoint and separation channels}
\label{appsubsec:midpoint_separation_channels}

Define
\begin{align}
\Delta\Sigma_{ab}^{(+)}(u)
&\equiv
\frac{1}{2}
\left[
\Delta\Sigma_{ab}^{[1]}(u)
+
\Delta\Sigma_{ab}^{[2]}(u)
\right],
\nonumber\\*
\Delta\Sigma_{ab}^{(-)}(u)
&\equiv
\frac{1}{2}
\left[
\Delta\Sigma_{ab}^{[1]}(u)
-
\Delta\Sigma_{ab}^{[2]}(u)
\right].
\label{eq:appb_symmetric_antisymmetric_scores}
\end{align}
Then
\begin{align}
\epsilon_1
\Delta\Sigma_{ab}^{[1]}
+
\epsilon_2
\Delta\Sigma_{ab}^{[2]}
={}&
(\epsilon_1+\epsilon_2)
\Delta\Sigma_{ab}^{(+)}
\nonumber\\*
&+
(\epsilon_1-\epsilon_2)
\Delta\Sigma_{ab}^{(-)}.
\label{eq:appb_endpoint_score_decomposition}
\end{align}
Equations~\eqref{eq:appb_endpoint_sum_difference} and \eqref{eq:appb_normalized_response_master} therefore give
\begin{align}
\delta_{\bm\beta}w_{ab}
={}&
\bm\beta\cdot\hat{\bm N}
\mathcal R_{ab}^{\rm mid}(\theta)
\nonumber\\*
&+
\bm\beta\cdot\hat{\bm s}
\mathcal R_{ab}^{\rm sep}(\theta),
\label{eq:appb_midpoint_separation_response}
\end{align}
where
\begin{align}
\mathcal R_{ab}^{\rm mid}(\theta)
={}&
2\cos\left(
\frac{\theta}{2}
\right)
\left[
1+w_{ab}(\theta)
\right]
\nonumber\\*
&\times
\left[
\mathcal G_{ab}(\theta)
+
\Delta\Sigma_{ab}^{(+)}(\theta)
\right],
\label{eq:appb_midpoint_coefficient}
\\
\mathcal R_{ab}^{\rm sep}(\theta)
={}&
2\sin\left(
\frac{\theta}{2}
\right)
\left[
1+w_{ab}(\theta)
\right]
\Delta\Sigma_{ab}^{(-)}(\theta).
\label{eq:appb_separation_coefficient}
\end{align}
For a same-tracer auto-correlation with identical endpoint selection and exchange-consistent pair completeness,
\begin{align}
\Delta\Sigma_{aa}^{[1]}(u)
=
\Delta\Sigma_{aa}^{[2]}(u),
\end{align}
and hence
\begin{align}
\mathcal R_{aa}^{\rm sep}(\theta)=0.
\label{eq:appb_auto_separation_response_zero}
\end{align}
For distinct tracers, $\Delta\Sigma_{ab}^{(-)}$ need not vanish. The separation channel is tracer antisymmetric:
\begin{align}
\mathcal R_{ba}^{\rm sep}(\theta)
=
-
\mathcal R_{ab}^{\rm sep}(\theta).
\label{eq:appb_cross_separation_antisymmetry}
\end{align}
Within the local angular correlation, a nonzero separation channel therefore measures a tracer-antisymmetric pair-conditioned residual. More general separation-odd observables constructed from explicit endpoint or mark ordering are described by the exchange-even total kernels of Sec.~\ref{subsec:marked_auto_pair_ordering}, rather than by an ordinary unordered auto-correlation.

The probabilistic interpretation of the residual selection terms as mark--clustering covariances is given in Sec.~\ref{subsec:mark_clustering_covariance}.

\subsection{Response of a finite angular bin}
\label{appsubsec:angular_bin_response}

The geometrical response may be represented either as a remapping of the pair density in observed coordinates or as a remapping of the pair kernel in rest-frame coordinates. To display the second representation, let
\begin{align}
K_B(u)
=
\Theta(u-u_-)
\Theta(u_+-u)
\label{eq:appb_top_hat_bin}
\end{align}
select $B=[u_-,u_+]$. At fixed rest-frame directions, the observed pair cosine is
\begin{align}
u
=
u'+
(1-u')
(\epsilon_1+\epsilon_2)
+
\mathcal O(\beta^2).
\label{eq:appb_inverse_pair_cosine_map}
\end{align}
The pulled-back kernel therefore varies as
\begin{align}
\delta_{\bm\beta}K_B
={}&
(1-u)
(\epsilon_1+\epsilon_2)
\frac{\partial K_B}{\partial u}
\nonumber\\*
={}&
(1-u)
(\epsilon_1+\epsilon_2)
\Bigl[
\deltadir(u-u_-)
-
\deltadir(u-u_+)
\Bigr].
\label{eq:appb_bin_boundary_flux}
\end{align}
Equation~\eqref{eq:appb_bin_boundary_flux} is the kernel part of the signed pair flux through the two bin boundaries. In observed coordinates the angular kernel is fixed, the parent density is remapped, and the angular Jacobians are included. In rest-frame coordinates the density and measure are fixed, while all observed angular factors, including selection masks, pair completeness, and weights, are transported. The equivalence is the Jacobian-plus-density identity in Eq.~\eqref{eq:angular_transport_identity}, not an equality of density remapping with the derivative of a single kernel in isolation. Do not add the two angular representations; retain mark and explicit analysis responses in whichever representation is chosen.

\subsection{Why there is no universal three-dimensional separation law}
\label{appsubsec:no_universal_3d_separation_law}

For a three-dimensional catalog, a pair separation depends on both angular and radial observables. Let $z_i$ denote the observer-frame redshift and let $\chi(z)$ be the radial coordinate adopted by the analysis. Define
\begin{align}
\bm x_i
&=
\chi(z_i)\hat{\bm n}_i,
\qquad
\bm r
=
\bm x_2-\bm x_1,
\nonumber\\*
r
&=
|\bm r|,
\qquad
\hat{\bm r}
=
\frac{\bm r}{r}.
\label{eq:appb_observed_3d_positions}
\end{align}
At fixed observed $(z_i,\hat{\bm n}_i)$, the corresponding cosmic-rest-frame redshift and direction satisfy
\begin{align}
1+z_i'
&=
\delta_i(1+z_i),
\nonumber\\*
\hat{\bm n}_i'
&=
\hat{\bm n}_i-
\bm\beta_{\perp i}
+
\mathcal O(\beta^2).
\label{eq:appb_3d_endpoint_transformations}
\end{align}
Define
\begin{align}
Q_{\chi,i}
\equiv
(1+z_i)
\frac{\partial\chi}{\partial z_i}.
\label{eq:appb_radial_response_coefficient}
\end{align}
The rest-frame position represented in the same radial convention is
\begin{align}
\bm x_i'
=
\bm x_i+
\epsilon_iQ_{\chi,i}\hat{\bm n}_i-
\chi_i\bm\beta_{\perp i}
+
\mathcal O(\beta^2),
\label{eq:appb_3d_position_response}
\end{align}
where $\chi_i\equiv\chi(z_i)$. With $\bm r'\equiv\bm x_2'-\bm x_1'$, the separation response is
\begin{align}
\delta_{\bm\beta}\ln r
={}&
\frac{1}{r}
\hat{\bm r}\cdot
\Bigl[
\epsilon_2Q_{\chi,2}\hat{\bm n}_2-
\epsilon_1Q_{\chi,1}\hat{\bm n}_1
\nonumber\\*
&\hspace{3.2em}-
\chi_2\bm\beta_{\perp2}+
\chi_1\bm\beta_{\perp1}
\Bigr].
\label{eq:appb_general_3d_separation_response}
\end{align}
The response depends on both endpoint distances, both projections of the observer velocity, the distance--redshift convention, and the full pair geometry. It does not reduce in general to a universal expression proportional to $\bm\beta\cdot\hat{\bm r}$. A three-dimensional analysis must additionally specify the radial selection, tracer ordering, and line-of-sight prescription. Source peculiar velocities and the other relativistic light-cone terms are separate additions to this observer-coordinate transformation. None of these complications alters the endpoint-exchange rule: with an exchange-even line of sight, an unordered same-tracer auto-correlation remains even under $\bm r\mapsto-\bm r$.

\subsection{Logarithmic-slope interpretation}
\label{appsubsec:logarithmic_slope_interpretation}

Equation~\eqref{eq:appb_pure_midpoint_response} may be written as
\begin{align}
\mathcal R_{\rm ab}^{\rm mid}(\theta)
=
\frac{
2\sin(\theta/2)
}{
\theta
}
\frac{\partial w_0}{\partial\ln\theta}.
\label{eq:appb_logarithmic_slope_form}
\end{align}
At small angular separation,
\begin{align}
\frac{
2\sin(\theta/2)
}{
\theta
}
=
1-
\frac{\theta^2}{24}
+
\mathcal O(\theta^4),
\end{align}
so the midpoint response approaches the logarithmic derivative of the angular correlation. This controls the shape of the boost template; it should not by itself be identified with the signal-to-noise ratio of a survey measurement.

For a localized feature
\begin{align}
 w_{\rm f}(\theta)
=
 A_{\rm f}
\exp\left[
-
\frac{
(\theta-\theta_{\rm f})^2
}{
2\sigma_{\rm f}^2
}
\right],
\end{align}
the geometrical response is
\begin{align}
\mathcal R_{{\rm f},{\rm ab}}^{\rm mid}(\theta)
=
-
2\sin\left(
\frac{\theta}{2}
\right)
\frac{
\theta-\theta_{\rm f}
}{
\sigma_{\rm f}^2
}
 w_{\rm f}(\theta).
\label{eq:appb_localized_feature_response}
\end{align}
This isolated-feature contribution vanishes at its center and changes sign across it. Its largest response is on the flanks, not at the feature peak. A smooth background correlation or a selection term need not vanish at that angle, so the statement does not apply automatically to the total measured template. A BAO-like feature can contribute spectral shape information, but separation of templates depends on their weighted linear independence. The harmonic transform of the same exchange-even response is derived in Sec.~\ref{sec:harmonic_coupling}.

\section{Harmonic Derivation and Normalization Checks}
\label{app:harmonic_details}

This Appendix fixes the phases and normalizations connecting the pair response to the quadratic coefficients. The harmonic convention is $a_{\ell m}=\int\Delta Y_{\ell m}^*\,\pd\Omega$ with orthonormal complex spherical harmonics and the usual Condon--Shortley phase.

\subsection{From endpoint modulation to the Gaunt kernel}

The addition theorem gives
\begin{align}
\mathcal F(u_{12})=\sum_{jm}F_jY_{jm}(\hat{\bm n}_1)Y_{jm}^*(\hat{\bm n}_2).
\end{align}
In a double harmonic transform of $\epsilon_1\mathcal F(u_{12})$, integration over endpoint 2 fixes $j=\ell_2$. The remaining endpoint integral is
\begin{align}
&\int Y_{\ell_1m_1}^*Y_{1M}Y_{\ell_2m_2}\,\pd\Omega\nonumber\\*
&\qquad=(-1)^{m_1}h_{\ell_1 1\ell_2}
\begin{pmatrix}\ell_1&1&\ell_2\\-m_1&M&m_2\end{pmatrix}.
\label{eq:app_gaunt_identity}
\end{align}
The $\epsilon_2$ term instead fixes $j=\ell_1$. Combining the plus and minus channels gives Eq.~\eqref{eq:general_harmonic_response_kernel}. The all-zero Wigner symbol excludes equal neighboring indices at first order because $\ell_1+\ell_2+1$ must be even.

The Clebsch--Gordan convention used in Eq.~\eqref{eq:sample_BipoSH_estimator} is
\begin{align}
&\langle\ell_1m_1\,\ell_2,-m_2\mid LM\rangle\nonumber\\*
&\quad=(-1)^{\ell_1-\ell_2+M}\sqrt{2L+1}
\begin{pmatrix}\ell_1&\ell_2&L\\m_1&-m_2&-M\end{pmatrix}.
\end{align}
Permuting the Wigner columns and using their orthogonality yields Eq.~\eqref{eq:biposh_template_definition}. For $\ell_2=\ell+1$, the phase is $(-1)^{-1}=-1$, so $\mathcal H_\ell=-h_{\ell\,1\,\ell+1}/\sqrt3$. Its square follows from the closed zero-row symbol and equals $(\ell+1)/(4\pi)$.

\subsection{Local-map mixing}

With $\bm\beta=\beta\hat{\bm z}$, the inverse-aberration operator is $-\nabla_S\epsilon\cdot\nabla_S=\beta\sin\vartheta\,\partial_\vartheta$. The identities
\begin{align}
\cos\vartheta\,Y_{\ell m}
&=G_{\ell+1,m}Y_{\ell+1,m}+G_{\ell m}Y_{\ell-1,m},\\
\sin\vartheta\,\partial_\vartheta Y_{\ell m}
&=\ell G_{\ell+1,m}Y_{\ell+1,m}-(\ell+1)G_{\ell m}Y_{\ell-1,m}
\end{align}
give, for a scalar signal modulation $\mathcal M$,
\begin{align}
\widetilde a_{\ell m}
={}&a_{\ell m}+\beta(\ell-1+\mathcal M)G_{\ell m}a_{\ell-1,m}
\nonumber\\*
&+\beta(\mathcal M-\ell-2)G_{\ell+1,m}a_{\ell+1,m}+\mathcal O(\beta^2).
\label{eq:app_local_map_harmonic_mixing}
\end{align}
Contracting with $a_{\ell+1,m}^*$ yields Eq.~\eqref{eq:auto_neighboring_covariance_local_map}. For a general response field $\mathcal U$, replace the two modulation contributions by $\beta G_{\ell m}u_{\ell-1,m}$ and $\beta G_{\ell+1,m}u_{\ell+1,m}$, obtaining the cross-spectra in Eq.~\eqref{eq:general_cross_spectral_response}. No discreteness transformation has been assumed in this signal calculation.

The configuration-space derivation gives an independent check. For $\mathcal F^{(+)}=-(1-u)w'(u)+\mathcal M w(u)$, integrate its Legendre coefficients against $P_\ell+P_{\ell+1}$. The endpoint term in integration by parts vanishes because $(1-u)(P_\ell+P_{\ell+1})$ is zero at both $u=1$ and $u=-1$, provided the regulated correlation is sufficiently regular. The identity
\begin{align}
\frac{\pd}{\pd u}\left[(1-u)(P_\ell+P_{\ell+1})\right]
=\ell P_\ell-(\ell+2)P_{\ell+1}
\end{align}
then gives
\begin{align}
F_\ell^{(+)}+F_{\ell+1}^{(+)}
={}&\ell C_\ell-(\ell+2)C_{\ell+1}\nonumber\\*
&+\mathcal M(C_\ell+C_{\ell+1}).
\label{eq:app_legendre_template_identity}
\end{align}
For a singular zero-separation correlation, finite resolution or an explicitly specified regularization is required before interchanging derivatives and integrals. The contact distribution is treated separately, as in Section~\ref{subsec:contact_covariance}.

\subsection{Gaussian null variance and its limitations}

Under the continuous Gaussian signal plus additive-noise model, Wick's theorem gives for $\ell_1\neq\ell_2$ a product $C_{\ell_1}^{\rm tot}C_{\ell_2}^{\rm tot}$ in the direct contraction. The swapped contraction connects the reversed multipole block; it is not a second copy in a data vector containing only $(\ell,\ell+1)$. Summing the squared Clebsch--Gordan coefficients gives unity and yields Eq.~\eqref{eq:BipoSH_estimator_variance}. For the auto $L=1$ neighboring block, $\widehat A^{1,-M}=(-1)^M\widehat A^{1M*}$; $M=0$ is real. A likelihood must use three independent real dipole components, not three unconstrained complex ones.

The full-sky expectation $\langle\widehat A\rangle=\mathcal T\epsilon$ and fixed covariance imply the estimator by minimizing the variance of a linear combination subject to unit response. Multiplying the weights by the true mean immediately gives the mismatch projection in Eq.~\eqref{eq:geometry_only_estimator_bias}. This derivation assumes a specified template and covariance. It does not justify ignoring non-Gaussian factorial moments, stochastic reference estimation, or residual contact means in a catalog measurement.

\section{Factorial Pair Covariance and Retained-Pair Sampling}
\label{app:factorial_covariance}

This Appendix makes explicit what is and is not fixed by a second-order response calculation. Consider one simple catalog with diffuse angular coordinates, factorial densities $\lambda^{(j)}$, and finite fourth factorial moments on the window. For symmetric real kernels define $U_K=\sum_{i\neq j}K(Z_i,Z_j)$ and $N_g=\sum_i g(Z_i)$. Products are expanded by whether two ordered pairs have four, three, or two distinct objects. The factorial-moment convention is that of \citet{2003Daley_point_processI,Last_Penrose_2017}.

\subsection{Exact contractions before additional pair retention}

Writing $K_{12}=K(Z_1,Z_2)$ and using $\pd Z_{1\cdots j}=\prod_{i=1}^j\pd Z_i$, the covariance is
\begin{align}
&\cov(U_K,U_H)\nonumber\\*
&\quad=\int K_{12}H_{34}\left[\lambda^{(4)}_{1234}-\lambda^{(2)}_{12}\lambda^{(2)}_{34}\right]\,\pd Z_{1234}\nonumber\\*
&\qquad+4\int K_{12}H_{13}\lambda^{(3)}_{123}\,\pd Z_{123}\nonumber\\*
&\qquad+2\int K_{12}H_{12}\lambda^{(2)}_{12}\,\pd Z_{12}.
\label{eq:factorial_U_covariance}
\end{align}
The factor four counts the choices of a shared endpoint. The factor two counts the identical and reversed ordered pair. The mixed and linear covariances are
\begin{align}
&\cov(U_K,N_g)\nonumber\\*
&\quad=\int K_{12}g_3\left[\lambda^{(3)}_{123}-\lambda^{(2)}_{12}\lambda_3\right]\,\pd Z_{123}\nonumber\\*
&\qquad+\int K_{12}(g_1+g_2)\lambda^{(2)}_{12}\,\pd Z_{12},
\label{eq:factorial_mixed_covariance}\\
&\cov(N_f,N_g)\nonumber\\*
&\quad=\int f_1g_2\left[\lambda^{(2)}_{12}-\lambda_1\lambda_2\right]\,\pd Z_{12}
+\int fg\lambda\,\pd Z.
\label{eq:factorial_linear_covariance}
\end{align}
These formulas follow by explicit coincidence counting and are not Gaussian approximations. Integrations over factorial densities refer to distinct object identities. Real and imaginary parts, or conjugation of the second kernel, give the corresponding complex formulas.

For a deterministic mean $\lambda$ and symmetric $K$, set $k_1(Z)=\int K(Z,Z')\lambda(Z')\,\pd Z'$. The centered factorial quadratic is
\begin{align}
Q_K=U_K-2N_{k_1}+M[K].
\label{eq:compensated_factorial_quadratic}
\end{align}
Its covariance is obtained from Eq.~\eqref{eq:factorial_U_covariance} by subtracting the two mixed terms with coefficients two and adding $4\cov(N_{k_1},N_{h_1})$. For a Poisson catalog all factorial densities factorize, and the shared-endpoint contractions cancel to give
\begin{align}
\cov(Q_K,Q_H)=2\int K_{12}H_{12}\lambda_1\lambda_2\,\pd Z_{12}.
\label{eq:poisson_compensated_covariance}
\end{align}
A factorial connected kernel also need not be positive semidefinite. For $N$ fixed independent uniform objects on a full sphere, $w=-1/N$, so its factorial monopole is $C_0=-4\pi/N$. The positive contact contribution $4\pi/N$ restores the exactly zero variance of the fixed total count. Thus a Gaussian signal interpretation of the factorial spectrum is an additional modeling assumption, not a consequence of removing self-pairs.

For a clustered catalog they do not generally cancel to a two-spectrum Wick expression. The bispectrum, trispectrum, and contact contractions can all matter. Deterministic normalization and self-pair subtraction specify the target mean; they do not provide those higher moments for free.

\subsection{Extra pair retention is a separate stochastic operation}

Suppose the object catalog is fixed as above and independent Bernoulli retention is subsequently applied to each unordered pair with probability $\mathcal C_{ij}$. The same retained pair list is used in $U_K^R$ and $U_H^R$. Conditional on the catalog, disjoint retention variables and those for different pairs sharing an endpoint are independent, but $R_{ij}^2=R_{ij}$. Therefore
\begin{align}
&\cov(U_K^R,U_H^R)\nonumber\\*
&\quad=\int K_{12}H_{34}\mathcal C_{12}\mathcal C_{34}\nonumber\\*
&\qquad\times\left[\lambda^{(4)}_{1234}-\lambda^{(2)}_{12}\lambda^{(2)}_{34}\right]\,\pd Z_{1234}\nonumber\\*
&\qquad+4\int K_{12}H_{13}\mathcal C_{12}\mathcal C_{13}\lambda^{(3)}_{123}\,\pd Z_{123}\nonumber\\*
&\qquad+2\int K_{12}H_{12}\mathcal C_{12}\lambda^{(2)}_{12}\,\pd Z_{12}.
\label{eq:retained_pair_covariance}
\end{align}
The last term contains $\mathcal C$, not $\mathcal C^2$. Merely replacing every kernel by its mean retained weight in a deterministic weighted-pair covariance would miss this term. For dependent pair-retention decisions, the products in the four- and three-object terms are replaced by the appropriate conditional joint retention probabilities, while the product of mean pair intensities is still subtracted. Thus an arbitrary pair-completeness response does not specify an exact estimator variance by itself.

As a simple check, for a homogeneous Poisson total count with mean $\mu$, $K=H=1$, and constant pair-retention probability $c$, the ordered retained-pair count has
\begin{align}
\mathbb E U_1^R=c\mu^2,\qquad
\operatorname{Var}(U_1^R)=4c^2\mu^3+2c\mu^2.
\label{eq:retained_poisson_count_check}
\end{align}
The same result follows by conditioning on the total count: $U_1^R=2\,\mathrm{Binomial}[N(N-1)/2,c]$. This provides an exact check on the coincidence factors. In a survey, corresponding calculations or simulations must include the actual detection and pair-retention mechanism, estimated references, masks, and any mark weights.

\end{appendix}

\section*{Data availability}

No observational data were used in this study. The deterministic numerical models, figure-generation code, machine-readable validation tables, and reproducibility documentation are archived in Zenodo as version v1.0.0 \citep{Takeuchi2026SelectionClusteringCode}, DOI: \url{https://doi.org/10.5281/zenodo.22858818}. The development repository is available at \url{https://github.com/TsutomuTakeuchiTTT/selection-clustering-coupling}. The software is released under the BSD-3-Clause license. No named-survey forecast or observational measurement is claimed.

\section*{Acknowledgements}
TTT thanks Sebastian von Hausegger for helpful insights.

This work was supported by JSPS Grants-in-Aid for Scientific Research (24H00247), JST CREST Grant No. JPMJCR24Q1, and the Collaboration Funding of The Institute of Statistical Mathematics “Machine-Learning-Based Cosmogony: From Structure Formation to Galaxy Evolution.”

TTT used OpenAI ChatGPT 5.6 Pro as a writing tool to help organize the manuscript, check parts of the mathematical formulation, and revise the language. The author specified the scientific assumptions, independently verified the derivations, references, and physical interpretations, reviewed and edited the content as needed, and takes full responsibility for the content of the article.

\bibliographystyle{aasjournal}
\bibliography{cosmic_dipole_OJAp}

\end{document}